\documentclass{aa}  
\usepackage{graphicx}
\usepackage{natbib}
\usepackage{txfonts}
\usepackage[colorlinks=true,citecolor=blue]{hyperref}
\usepackage{xcolor}

\begin{document}

\title{Angular momentum transport by magnetic fields in main sequence stars with Gamma Doradus pulsators}

      \author{F.D. Moyano \inst{1}
        \and
        P. Eggenberger \inst{1}
        \and
        S.J.A.J. Salmon\inst{1}
        \and        
        J. S. G. Mombarg \inst{2}
        \and
        S. Ekstr\"om \inst{1}        
      }
          
      \institute{Observatoire de Gen\`eve, Universit\'e de Gen\`eve, 51 Ch. Pegasi, CH-1290 Versoix, Suisse
        \\ email: facundo.moyano@unige.ch
        \and Universit\'e de Toulouse, CNRS, UPS, CNES, 14 avenue \'Edouard Belin, F-31400 Toulouse, France
      }
          \date{Received --; accepted 30 June 2023}
          \titlerunning{Angular momentum transport by internal magnetic fields in main sequence stars}

 
  \abstract
      {Asteroseismic studies showed that cores of post main-sequence stars rotate slower than theoretically predicted by stellar models with purely hydrodynamical transport processes.
      Recent studies on main sequence stars, particularly Gamma Doradus ($\gamma$ Dor) stars, revealed their internal rotation rate for hundreds of stars, offering a counterpart on the main sequence for studies of angular momentum transport.}
      {We investigate whether such a disagreement between observed and predicted internal rotation rates is present in main sequence stars by studying angular momentum transport in $\gamma$ Dor stars.
Furthermore, we test whether models of rotating stars with internal magnetic fields can reproduce their rotational properties.}
      {We compute rotating models with the Geneva stellar evolution code taking into account meridional circulation and the shear instability.
       We also compute models with internal magnetic fields using a general formalism for transport by the Tayler-Spruit dynamo.
      We then compare these models to observational constraints for $\gamma$ Dor stars that we compiled from the literature, combining so the core rotation rates, projected rotational velocities from spectroscopy, and constraints on their fundamental parameters.}
      {We show that combining the different observational constraints available for $\gamma$ Dor stars enable to clearly distinguish the different scenarios for internal angular momentum transport.
      Stellar models with purely hydrodynamical processes are in disagreement with the data whereas models with internal magnetic fields can reproduce both core and surface constraints simultaneously.}
   {Similarly to results obtained for subgiant and red giant stars, angular momentum transport in radiative regions of $\gamma$ Dor stars is highly efficient, in good agreement with predictions of models with internal magnetic fields.}

   \keywords{asteroseismology --
             stars: rotation --
             stars: interiors  --
             stars: evolution --
             stars: variables: Gamma Doradus --
             methods: numerical}

   \maketitle
%

   \section{Introduction}
   \label{intro}   
   Main sequence (MS) stars are some of the most studied objects in stellar physics.
   They are numerous, long-lived and can be found in all types of stellar systems, from binary stars to galaxies.
   They represent the longest evolutionary phase until they end their lives and become either compact objects, or leave no remnant behind, depending on their initial masses.
   Their accurate characterisation let us infer fundamental information like ages, masses, and chemical compositions, which are of great importance to other fields from exoplanetary studies to galactic archaeology.
   This in turn requires accurate stellar structure and evolution models.
   However despite our detailed understanding of MS stars there is still plenty of details in current debate about their structure.
   One current issue in our understanding of stellar interiors in MS stars is the transport of angular momentum (AM).
     An accurate physical description of AM transport in stellar interiors is crucial to make predictions based on stellar models due to the transport of chemical elements arising from rotational instabilities; this is particularly relevant for massive stars \citep[e.g.][]{maeder00}.
     Other long-standing problems remain unsolved nowadays, like the convective overshooting, which is thought to be due to the penetration of convective matter beyond the limits imposed by classical convection criteria and is necessary to explain several features from low- to high-mass stars \citep[e.g.][]{claret19,martinet21}.
   Due to the complex interplay between AM transport, rotational mixing, and other uncertainties such as convective overshooting, it is fundamental to characterize the AM redistribution in stellar interiors.
   
   Thanks to the continuous developments of asteroseismology it is now possible to obtain information about stellar interiors and in particular measurements of internal rotation rates \citep[e.g.][]{aerts21}.
   Indeed, thanks to space-borne missions, detailed asteroseismic studies were made possible for a large number of stars in different evolutionary phases and mass-ranges leading to the measurement of rotation rates in stellar interiors \citep{aerts03,mosser12,deheuvels12,deheuvels14,deheuvels15,vanreeth16,papics17,gehan18,tayar19,ouazzani19,deheuvels20,li20,garcia22}.
   These space missions include \textit{CoRoT} \citep{baglin09}, \textit{Kepler} \citep{borucki10}, \textit{TESS} \citep{ricker15}, and in the future \textit{PLATO} \citep{rauer14}, and provided us with long and continuous photometry which made possible the characterisation of several pulsation modes for single stars and unprecedented frequency resolution.
   This would otherwise be difficult to obtain from ground-based observations.
   In particular for MS stars this made possible the exploitation of gravity modes, which are known to have relatively long periods as compared to pressure modes.
   These modes are mainly sensitive to the inner regions of MS stars and thus can give us critical information about their internal structure.

   Regarding the theoretical modelling of AM transport in stellar interiors, a first hint on an improper treatment was evidenced by the solar rotation profile.
   Hydrodynamical processes transport insufficient AM and cannot account for it \citep{pinsonneault89,chaboyer95, charbonnel05, eggenberger05}.
   This challenges the view of AM transport driven solely by meridional currents and hydrodynamical turbulent processes \citep{zahn92}, which were employed in extensive grids of stellar models with rotation \citep[e.g.][]{ekstrom12}.
   A similar conclusion was obtained for evolved stars, in particular for red giants \citep{eggenberger12,marques13,ceillier13,moyano22}.
   As a consequence, several studies were done to identify an additional process able to transport AM more efficiently and assess its impact on the internal AM redistribution.
   Some of these processes include small-scale internal magnetic fields \citep{spruit02, cantiello14, fuller19, eggenberger22b}, internal gravity waves \citep[e.g.][]{charbonnel05, fuller14, pincon17}, transport by mixed modes \citep{belkacem15a,belkacem15b}, pumping of AM from convective envelopes \citep{kissin15}, magneto-rotational instabilities \citep[e.g.][]{wheeler15, spada16, griffiths22,moyano23}, among others.
   However, despite strong efforts none of them gives yet a satisfactory answer, pointing out to a general misunderstanding of the problem.
   
   Among MS stars, one interesting group of stars for AM studies are Gamma Doradus ($\gamma$ Dor) stars, which are variable late-F to early-A type stars.
  Their pulsations periods are typically $P \simeq 1$ day but can range from $P \simeq 0.3$ to $P \simeq 3$ days and correspond to high-order non-radial gravity modes.
   They are found at effective temperatures of $T_{\rm eff} \simeq 6700 - 7900$ K, surface gravities of $\log g \simeq 3.9 - 4.3$ dex and  metallicities slightly sub-solar, and belong mainly to the MS \citep{kaye99,tkachenko13}.
   Their masses are $M \simeq 1.4 - 2.0 M_{\odot}$ and are thus expected to have a convective core surrounded by an extensive radiative region and a shallow convective envelope.
   
   Gravity modes (g-modes) are sensitive to the properties of the regions close to the stellar core.
   For a non-rotating homogeneous star with a convective core and a radiative envelope high-radial order gravity modes of same angular degree are expected to be equally spaced in period \citep{tassoul80}.
   In particular \citet{miglio08} showed how the internal stratification of chemicals and entropy can lead to differences in the period spacing ($\Delta P$), defined as the difference in period between modes of consecutive radial orders but same angular and azimuthal degrees.
   \citet{bouabid13} further studied the effects of rotation and chemical mixing, showing that the period spacing changes with the pulsation period when rotation is taken into account.    
   This was further pursued by \citet{ouazzani17} who showed how rotation can induce a slope in the period spacing pattern (defined as the relation between the pulsation periods and the period spacing of subsequent radial modes of same angular degree) of g-modes and showed that it is mainly sensitive to the rotation rate of the star.
   Furthermore, \citet{christophe18} developed a method based on the traditional approximation of rotation with which the near-core rotation rate can be obtained by exploiting the properties of high-order g-modes.
   This methodology was then applied to a sample of 37 stars by \citet{ouazzani19} who then obtained their near-core rotation rates \citep[see also][]{vanreeth16}.
   They further studied the internal AM transport due to hydrodynamical processes under different assumptions, concluding that a more efficient process should be present in stellar interiors of MS stars.
   Later on, \citet{li20} studied a larger sample of $\gamma$ Dor stars and obtained their near-core rotation rates for 611 of them, among other asteroseismic parameters.
   However, it remains unclear how efficient during the MS the AM transport should be, and if it is consistent with constraints obtained for other stars, in particular regarding transport by the magnetic Tayler instability.

   In this paper we study the role of AM transport by internal magnetic fields in MS stars taking advantage of large sample of $\gamma$ Dor stars  with well characterised rotational constraints.
   In particular we combine internal rotation rates measured with asteroseismic techniques, and surface rotational velocities from spectroscopic studies.

  The paper is organised as follows: in Sect. \ref{data} we describe the data collected from the literature and in Sect. \ref{methods} we describe the input physics used in our models and the indicators used to compare with observations.
  In Sect. \ref{results} we present the various comparisons made with observational data, and discuss the uncertainties present in models that could impact our conclusions in Sect. \ref{uncertain}.
  We provide additional evidence to our analysis in Sect. \ref{additional_evidence}, then discuss our results in the context of more evolved stars in Sect. \ref{discussion} and we finally summarise and conclude in Sect \ref{conclusion}.

   \section{Data: fundamental parameters, core rotation rates and surface velocities}
   \label{data}
\begin{figure}[htb!]
     \resizebox{\hsize}{!}{\includegraphics{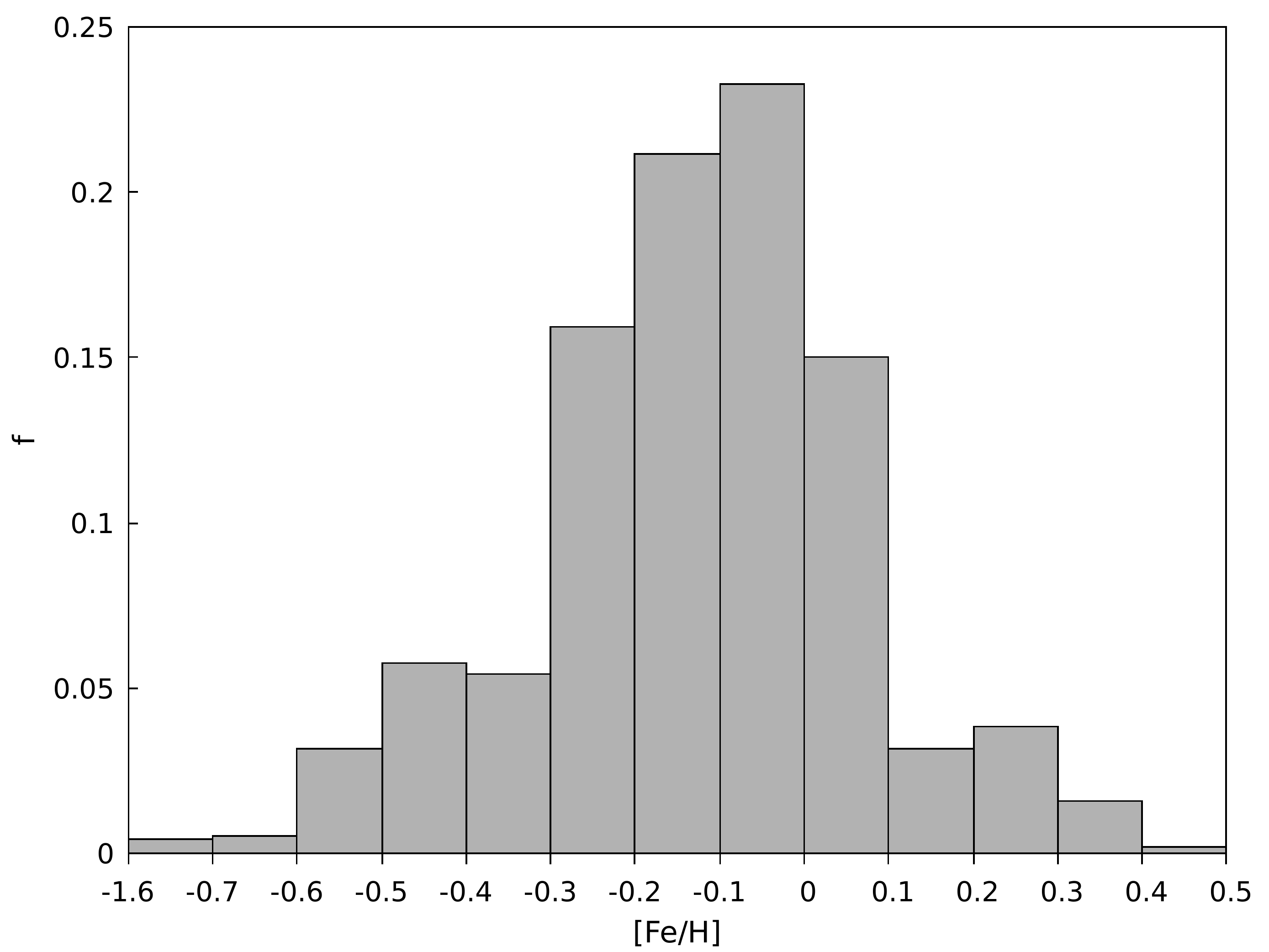}}
     \caption{Normalised distribution of metallicities of $\gamma$ Dor stars used in this work.
       The histogram contains 572 stars with metallicities from \citet{mathur17}.
       The leftmost bin contains all the outliers with [Fe/H] between -1.6 and -0.7 dex.
       The mean value of the sample is $\overline{\rm [Fe/H]} = -0.127 \pm 0.191$ dex (after removing the outliers).}
     \label{histogram_feh}
   \end{figure}

   Our study of internal AM redistribution mainly relies on the sample of $\gamma$ Dor stars provided by \citet{li20}, which contains 611 stars with measurements of their core rotation rate.
   They also provide measurements of the buoyancy radius which we use as a proxy for the evolution during the MS (see Sect. \ref{evol_indicator}).
   In addition to this, since the whole sample of stars lies in the \textit{Kepler} field, the effective temperature and metallicities are available from the \textit{Kepler} catalogue of revised stellar properties \citep{mathur17}.
   The distribution of metallicities is shown in Fig. \ref{histogram_feh}, its mean value is ${\rm [Fe/H]} = -0.127$ dex with a standard deviation of $\sigma_{\rm [Fe/H]}=0.191$ dex.
   Approximately $75 \%$ of our stars have metallicities between ${\rm[Fe/H]}= -0.3$ and ${\rm [Fe/H]}= -0.1$ dex (see Fig. \ref{histogram_feh}).
   We note that we adopt a slightly lower metallicity than the one adopted previously by \citet{ouazzani19} who used ${\rm[M/H]} = 0.11 \pm 0.17$ dex.
   The cause of this difference was discussed by \citet{gebruers21} and is likely due to issues related to the continuum fit in the sample of \citet{vanreeth16} which was used by \citet{ouazzani19}.
   However, this difference in metallicity does not have a major impact on the conclusions regarding AM transport.
    %
    
\begin{figure}[htb!]
     \resizebox{\hsize}{!}{\includegraphics{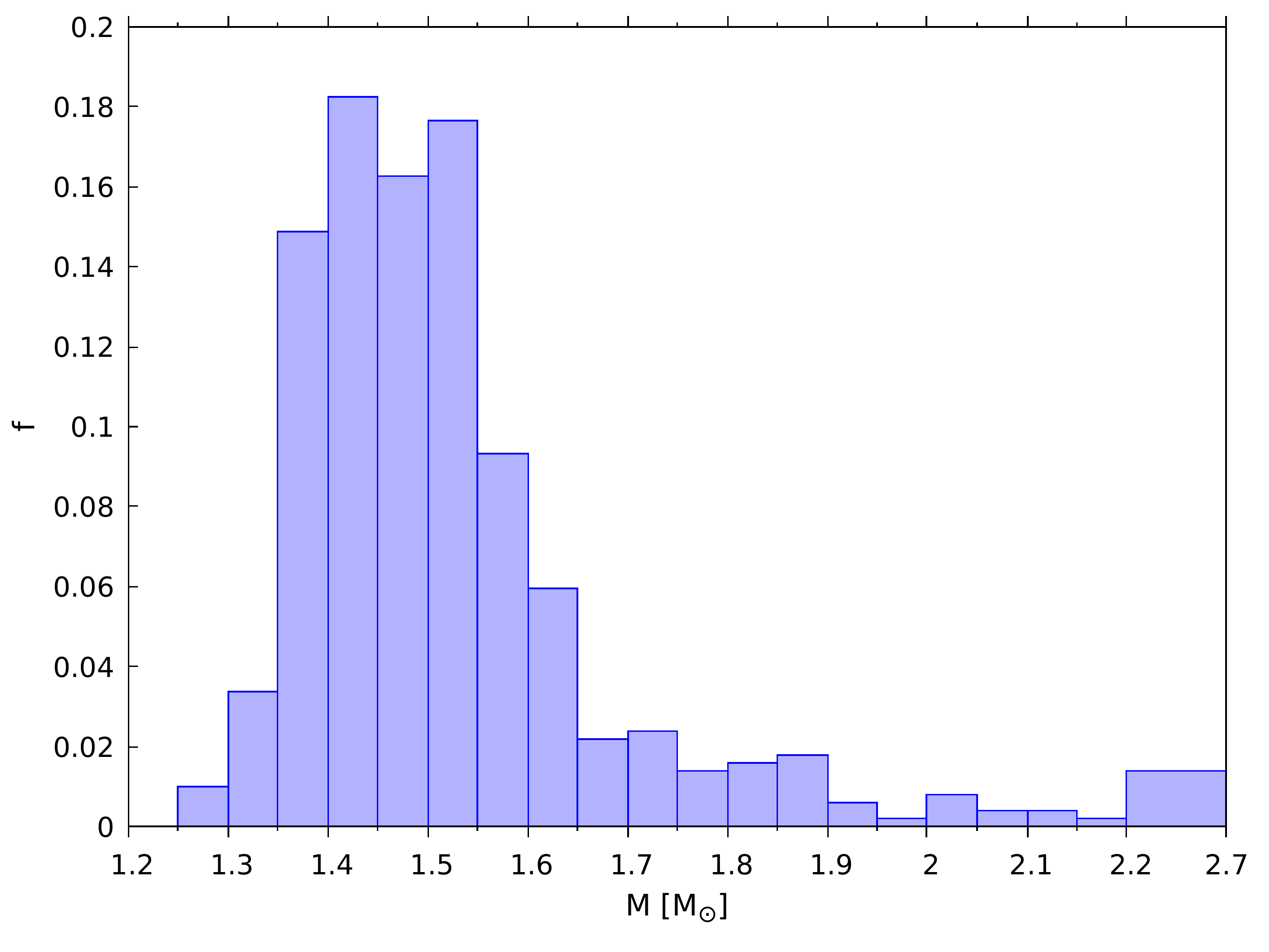}}
     \caption{Normalised distribution of stellar masses of $\gamma$ Dor stars used in this work.
       The histogram contains 504 stars with masses determined by \citet{murphy19}.
       The rightmost bin contains all the outliers with masses between 2.2 and 2.7 $M_{\odot}$.
     The mean value of the sample is $\overline{\rm M}=1.51 \pm 0.14 M_{\odot}$ (after removing the outliers).}
     \label{histogram_mass}
   \end{figure}

\begin{figure}[htb!]
     \resizebox{\hsize}{!}{\includegraphics{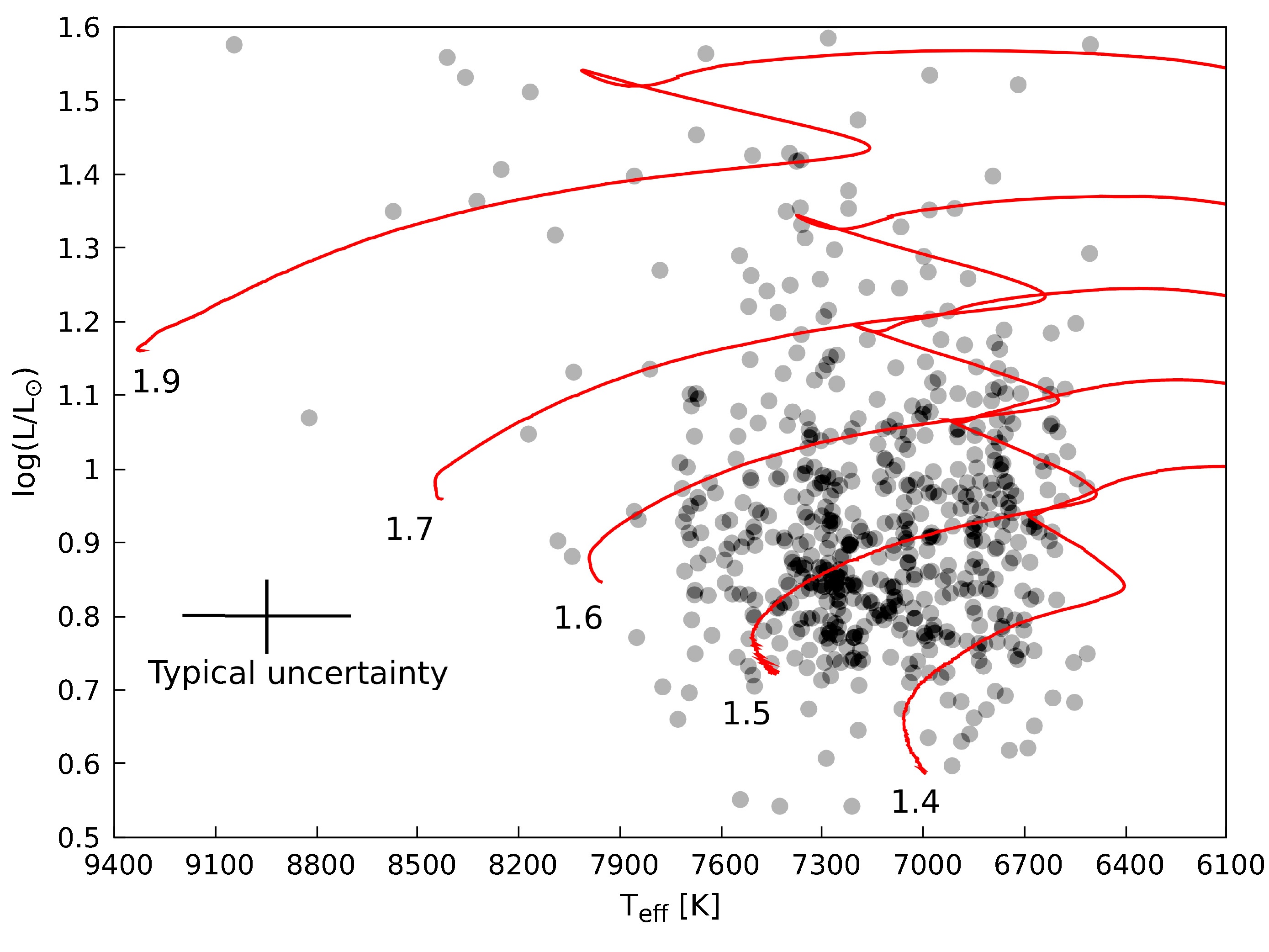}}
     \caption{Hertzsprung-Russell diagram showing a set of our models along with the data.
       The data points correspond to the $\gamma$ Dor stars studied in this work from \citet{li20},
       points are semitransparent to illustrate the number density when two or more stars overlap.
       The typical uncertainty on both effective temperature and luminosity is shown in the figure.
       The stellar models are computed from the ZAMS with an initial metallicity of $Z=0.01$ and an initial rotation rate $\Omega/2\pi = 20 \mu\text{Hz}$ (no magnetic fields); initial masses are indicated in solar masses for each track.
     }
     \label{hr_diagram}
   \end{figure}

   In addition to the constraints mentioned above, the bolometric luminosities for nearly all of the  $\gamma$ Dor stars in our sample were determined by \citet{murphy19}, who rederived the distances for nearly all of them (575 out of 611) based on the parallaxes from GAIA DR2 \citep{gaia16}.
   They also gave an estimate of the stellar masses for 575 stars in our sample by comparing stellar evolutionary tracks to the data based on the effective temperatures, bolometric luminosities and metallicities.
   This enables us to have a distribution of the stellar masses which we show in Fig. \ref{histogram_mass}, and thus we can make an accurate choice on the initial masses of our models.
   A Hertzsprung-Russell (HR) diagram comparing our models with rotation computed at $Z=0.01$ and the observational data is shown in Fig. \ref{hr_diagram}.
   We note that although the expected mass-range of $\gamma$ Dor stars is $M=1.4 - 2.0 M_{\odot}$, a few stars have inferred masses higher than $M=2.2 M_{\odot}$.
   This could probably occur because of the single-metallicity stellar tracks used by \citet{murphy19} and the fact that some stars have quite low metallicities (see Fig. \ref{histogram_feh}), which would mimic a star of lower metallicity but higher mass in the HR diagram, among other possible effects.
   After removing these stars the mean value of the stellar mass averaged over the whole sample is $M=1.51 \pm 0.14 M_{\odot}$.
   %
   
   To obtain more information on the AM content of our sample of $\gamma$ Dor stars, we also made a compilation of projected surface rotational velocities ($V \sin i$), determined from medium- to high-resolution spectra from the literature.
   This allows us to have information on the AM content both in the core and at the surface for a subsample of our stars, which is a fundamental point in our study.
   Our subsample of stars with determined surface velocities mainly comes from the study of \citet{gebruers21}, who determined $V \sin i$ for 88 of our stars with high-resolution spectra taken with the HERMES spectrograph \citep{raskin11} and whose resolution reaches $R \simeq 85000$.
   The rest of the stars come from either individual studies or cross-matches with surveys.
   These include:
   \begin{itemize}
\item  Three stars from \citet{vanreeth15b}: KIC5254203, KIC6468987, and KIC9480469.
\item  Three stars from individual studies:  KIC8975515 \citep{samadi-ghadim20}, and KIC10080943A,B \citep{schmid15,schmid16}.
\item  Two stars from LAMOST-DR7-MRS \citep{liu20,luo22}: KIC8495755 and KIC7448050.
\item  Five stars from \citet{sun21}: KIC6780397, KIC7761855, KIC7202395, KIC8759258, and KIC9598448.
\item  Three stars from \citet{frasca22}: KIC9348946, KIC6366512, and KIC7436266.
   \end{itemize}
    The stars studied by \citet{luo22, sun21, frasca22} come from medium resolution spectra with a spectral resolution of $R \simeq 7500$, namely the LAMOST-MRS \footnote{\url{http://dr7.lamost.org/v2.0/}} \citep{liu20}.
    For the surface velocities determined from medium-resolution spectra we only choose those stars who were observed with a signal-to-noise ratio higher than one hundred (S/N > 100).
    But although KIC9598448 has a $\textrm{S/N} = 98.6$, we decided to include it in our sample.
   When there is an overlapping star in any of these surveys with the sample of \citet{gebruers21}, we decided to use the parameters given by these authors due to the higher resolution of the spectra used in their work.
   Thus our subsample of stars for which we have both core rotation rates and projected surface rotational velocities contains 104 stars.

      We emphasise that the measurements of both quantities were obtained with completely different techniques and that we do not make any assumption on any of these quantities.
      \section{Physical ingredients of stellar models and asteroseismic observables}
      \label{methods}
    We compute all of our models with the Geneva stellar evolution code \citep[{\fontfamily{qcr}\selectfont GENEC};][]{eggenberger08}.
    This code was extensively used to compute grids of models of rotating stars in different mass ranges and metallicities \citep{ekstrom12, georgy13, eggenberger21, yusof22}.
    Here we mention only the relevant initial conditions and equations for the treatment of AM transport and we refer the reader to the mentioned works for a detailed description of the code and usual input physics adopted.

    The mean value of the metallicity distribution given by the data (see Fig. \ref{histogram_feh}) is $\overline{\rm [Fe/H]} = -0.127$, with a standard deviation of $\sigma_{\rm [Fe/H]} =0.191$ dex, which corresponds to $Z \simeq 0.0106 \pm 0.0040$ in our models.
    We thus choose an initial metallicity of $Z=0.01$ for our models, adopting the chemical mixture of heavy elements from \citet{asplund09}.
    We compute the evolution of all of our models from the zero-age main sequence (ZAMS) where we assume constant angular velocity along the interior as an initial condition.
    Initial velocities are chosen to reproduce the range in core rotation rates observed for $\gamma$ Dor stars at high buoyancy radius that should then be young MS stars.
    The values chosen are $\Omega/2\pi = 10, 20, {\rm and}$ $30 \mu{\rm Hz}$, which correspond to surface velocities of $V \simeq 60, 120$ {\rm and} 200 km/s for our $1.5 M_{\odot}$ models at $Z=0.01$.
    These velocities correspond to  16, 32, and 50 \%, respectively, of their critical velocities at the ZAMS (i.e. the rotational velocities at which the centrifugal force due to rotation counteracts the surface gravity).
    We adopt an extension of the convective core by the step overshooting formalism (i.e. the length over which overshooting acts is assumed to mix instantaneously), and so we choose to extend the convective core by 0.05 times the local pressure scale-height $H_{\rm p}$ for masses $1.25 \le M/M_{\odot} \le 1.7$ and $r=0.1 H_{\rm p}$ for masses $M > 1.7 M_{\odot} $.

    \subsection{AM transport: hydrodynamical processes and internal magnetic fields}
    To follow the redistribution of internal AM, we obtain the angular velocity at each layer of the stellar interior under the shellular approximation \citep{zahn92}.
    While in convective zones we assume uniform rotation, in radiative zones the AM transport is treated in an advecto-diffusive way \citep{zahn92}, by solving the equation
  \begin{equation}
  \rho \frac{{\rm d}}{{\rm d}t} \left( r^{2}\Omega \right)_{M_r}
  =  \frac{1}{5r^{2}}\frac{\partial }{\partial r} \left(\rho r^{4}\Omega
  U_{\rm r}\right)
  + \frac{1}{r^{2}}\frac{\partial }{\partial r}\left(\rho D r^{4}
  \frac{\partial \Omega}{\partial r} \right) \
\label{eq_amt_genec}
\end{equation}
  where $r$ is the radial coordinate, $\rho$ is the density, $\Omega$ is the horizontally-averaged angular velocity, $U_{\rm r}$ is the radial component of the meridional circulation velocity, and $D$ is the diffusion coefficient of AM.
  In our work, we consider two types of models: purely hydrodynamical and magneto-hydrodynamical.
  The former include only hydrodynamical transport of both AM and chemicals, specifically driven by the shear instability and meridional currents \citep{zahn92}.
  The latter include the hydrodynamical processes but also the transport by internal magnetic fields following the formalism of the Tayler-Spruit (TS) dynamo \citep{spruit02} in its calibrated version to reproduce asteroseismic rotation rates of evolved stars \citep{eggenberger22b}.
  
  For models without internal magnetic fields, in addition to advective transport by meridional circulation, the diffusive transport by the shear instability is taken into account and the diffusion coefficient is then $D=D_{\rm shear}$, following \citet{maeder97}.
  For the models with internal magnetic fields, the transport of AM is implemented as an additional source of viscosity which we compute as \citep{eggenberger22b}
  \begin{equation}
    \nu_{\rm mag}= C_{\rm T}^3 r^2 \Omega q^2 \left( \frac{\Omega}{N_{\rm eff}} \right)^2
    \end{equation}
  where $q \equiv \partial \rm{ln}\Omega / \partial \rm{ln} r$ is the local shear, $N_{\rm eff}$ the effective Brunt-V\"ais\"al\"a frequency, defined as $N^2_{\rm eff} = \frac{\eta}{K} N^2_{\rm T} + N^2_{\mu}$ with $\eta$ the magnetic diffusivity, $K$ the thermal diffisuvity, and $N_{\rm T}$ and $N_{\mu}$ the thermal and chemical components.
  Finally $C_{\rm T} = 216$ is a constant which was calibrated to reproduce the core rotation rate of evolved stars \citep[see][]{eggenberger22b}.
  This viscosity is added to the diffusion coefficient, hence the total diffusion coefficient of AM transport is $D=D_{\rm shear} + \nu_{\rm mag}$ in our models with internal magnetic fields.
  The magnetic instability is triggered when the local shear is above a critical value given by
  \begin{equation}
    q_{\rm min} = C_{\rm T}^{-1} \left( \frac{N_{\rm eff}}{\Omega} \right)^{7/4} \left( \frac{\eta}{r^2 N_{\rm eff}} \right) ^{1/4}.
  \end{equation}

  \begin{figure*}[ht!]
    \centering
     \resizebox{\hsize}{!}{\includegraphics[width=17cm]{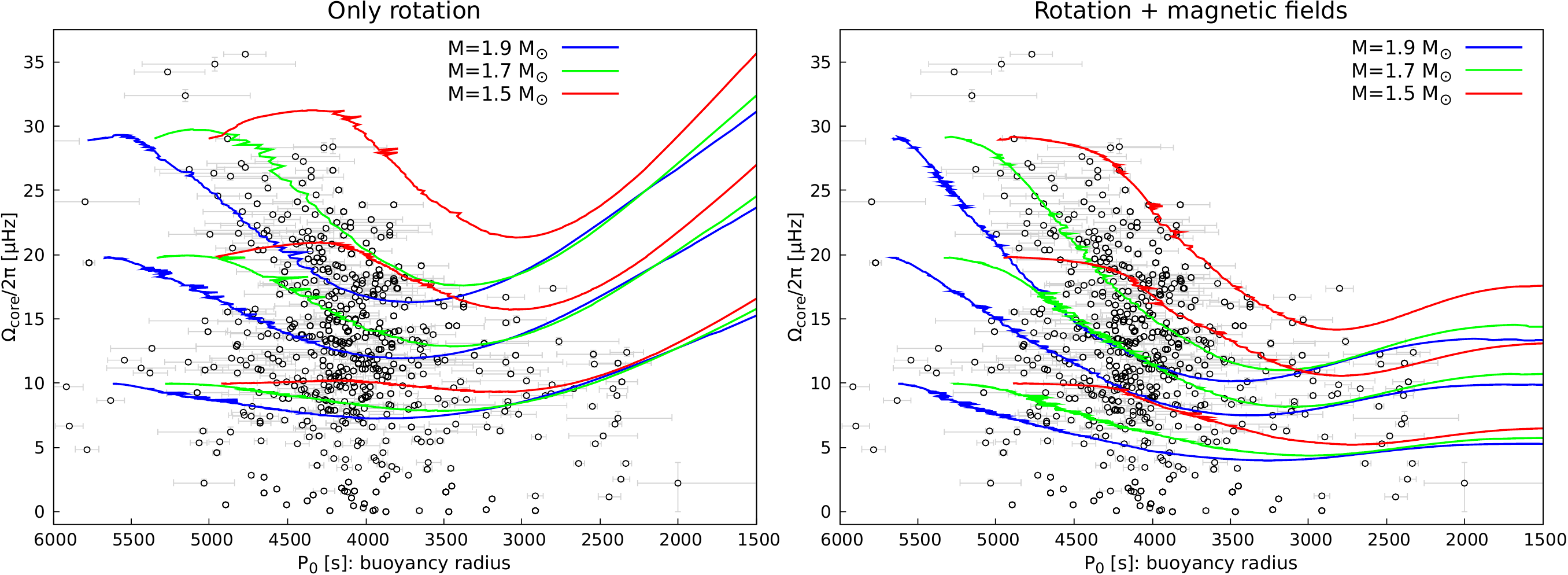}}
     \caption{Core rotation rate as a function of the buoyancy radius for models with different initial masses.
       \textbf{\textit{Left}}: AM transport driven by hydrodynamical processes alone.
       \textbf{\textit{Right}}: including also internal magnetic fields.
       The data points correspond to the $\gamma$ Dor stars presented by \citet{li20}.
       For each initial mass we show models starting at an initial rotation rate of $\Omega_{\rm core}/2\pi = 10, 20$ and $30 \mu$Hz.
       Models are computed from the ZAMS and the evolution goes from left to right.
     }
     \label{omegac_pi0_2cols}
   \end{figure*}
    \begin{figure*}[h!]
    \centering
     \resizebox{\hsize}{!}{\includegraphics[width=17cm]{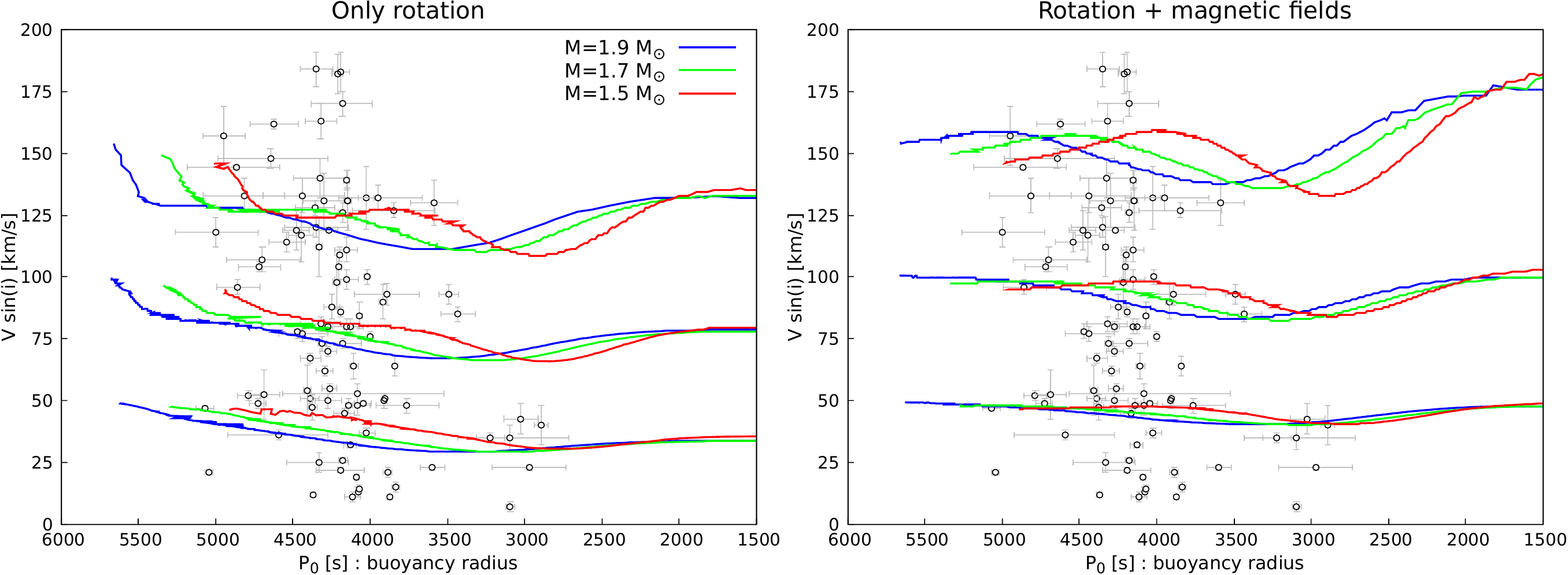}}
     \caption{Surface rotational velocity as a function of the buoyancy radius.
       The lines correspond to our models with different initial masses, computed at three different initial rotation rates for each initial mass.
       \textbf{\textit{Left}}:  including only hydrodynamical processes.
       \textbf{\textit{Right}}: including also internal magnetic fields.
       The data points correspond to $\gamma$ Dor stars for which surface velocities are available from the literature (see Sect. \ref{data}); buoyancy radii are taken from \citet{li20}.
       The surface velocity of the models is multiplied by $\pi/4$ to average over the unknown inclination angles of the stars.}
     \label{vsini_pi0_2cols}
   \end{figure*}
    %

  \subsection{Evolutionary indicator during the main sequence}
  \label{evol_indicator}
   Similarly to \citet{ouazzani19}, we use the buoyancy radius $P_{0}$ as an evolutionary indicator during the MS.
   We compute the buoyancy radius in our stellar models as
  \begin{equation}
    P_{0} = 2\pi^2 \left(\int_{r_{\rm in}}^{r_{\rm out}} N \frac{dr}{r}\right)^{-1}
  \end{equation}
  where $N$ is the Brunt-V\"ais\"al\"a frequency, $r$ is the radial coordinate, and $r_{\rm in}$ and $r_{\rm out}$ are the lower and upper limits of the g-mode cavity, respectively.
  We define the g-mode cavity as the region where the frequency of a high radial order mode, which we take as $k=-25$, is lower than both the Brunt-V\"ais\"al\"a and Lamb frequencies for dipole modes.
  
  As shown by \citet{ouazzani19}, the buoyancy radius decreases during the MS in the mass range $M=1.4 - 1.8 M_{\odot}$, suitable for $\gamma$ Dor stars (see their Fig. 2).
  This is because it depends mainly on the size of the convective core and both thermal and chemical stratification in the radiative layers just above it, which change through evolution because of the hydrogen burning and either expansion or contraction of the inner radiative regions.
  This can modify the value of the buoyancy radius, because it changes the chemical stratification in the near-core regions, specially if the core is retreating.
  This can be seen by writing the Brunt-V\"ais\"al\"a frequency as
  \begin{equation}
    \label{eq_bvfreq}
    N^2= \frac{g \delta}{H_{\rm p}} \left(\nabla_{\rm ad} - \nabla_{\rm rad} + \frac{\phi}{\delta} \nabla_{\mu} \right)
  \end{equation}
  where $\nabla_{\mu} \equiv \partial {\rm \ln \mu} / \partial {\rm \ln P}$ is the value of the chemical gradient, $\nabla_{\rm rad}$ is the temperature gradient, $\nabla_{\rm ad}$ is the adiabatic temperature gradient, $g$ is the local gravity, and $\phi \equiv (\partial \ln \rho/ \partial \ln \mu)_{P,T}$ and $\delta \equiv -(\partial\ln \rho / \partial\ln T)_{P,\mu}$.
  As shown by Eq. \ref{eq_bvfreq}, the Brunt-V\"ais\"al\"a frequency depends on both the entropy stratification, via the thermal and adiabatic gradients, and the chemical stratification via the chemical gradient.
 We use the buoyancy radius as an evolutionary indicator during the MS, taking advantage of the measurements provided by \citet{li20}.

    \subsection{Rotation rate in the gravity-mode cavity}
    The core rotation rates used in this work are not that of the convective core, but rather an average rotation rate over the g-mode cavity (usually denoted near-core); to compare our models with the data we then compute the near-core rotation rate from our models as \citep{ouazzani19}
  \begin{equation}
    \label{eq_meanomega}
    \Omega_{\rm core}= \frac{\int_{r_{\rm in}}^{r_{\rm out}}{\Omega N dr/r}}{\int_{r_{\rm in}}^{r_{\rm out}}{N dr/r}}
  \end{equation}
  where the limits of the integrals are the same as for the buoyancy radius and denote the g-mode cavity.
  This variable represents the near-core rotation rate as sensed by gravity modes which we use to compare to the rotation rates available in our sample.
  In the following sections we refer to it just as the core rotation rate.

    \begin{figure*}[ht!]
    \centering
     \resizebox{\hsize}{!}{\includegraphics[width=17cm]{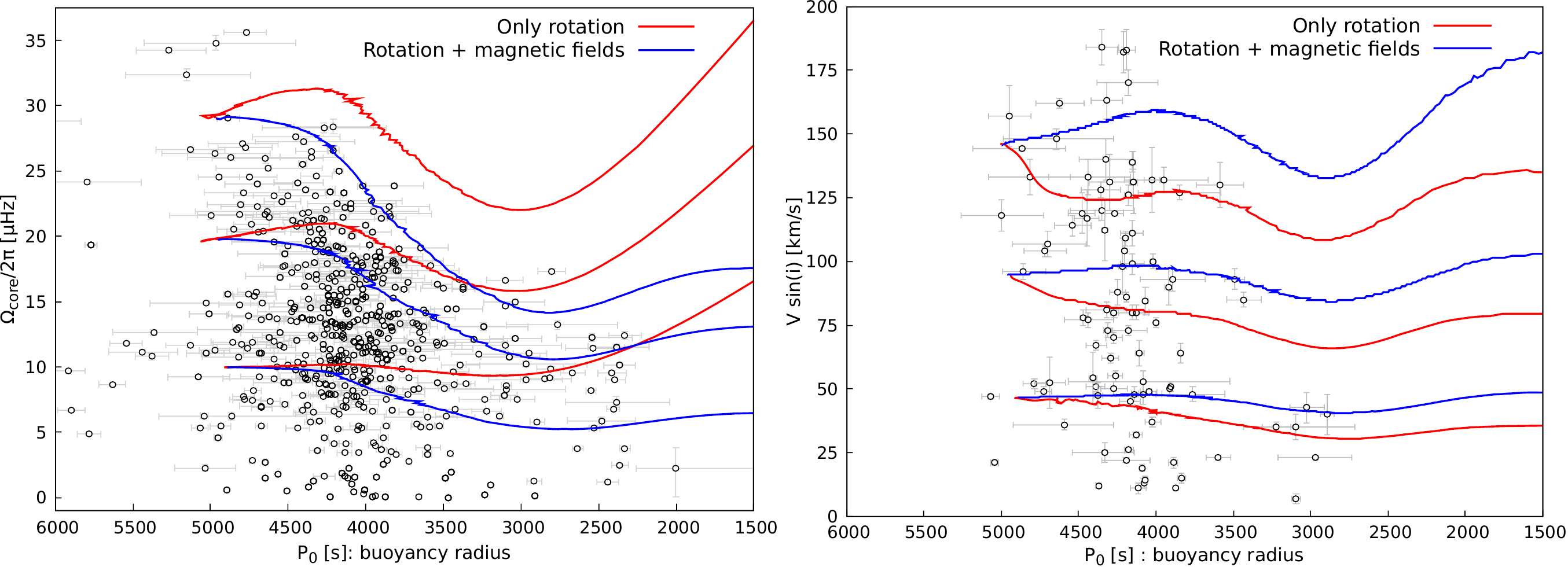}}
     \caption{Effect of internal magnetic fields on the redistribution of AM during the MS; evolution goes from left to right.
       \textbf{\emph{Left:}} Core rotation rate.
       \textbf{\emph{Right:}} Surface rotational velocity.
       The surface velocity of the models are multiplied by a factor $\pi/4$ to average over the unknown inclination angles of the sample.
       The models have an initial mass and metallicity of $M=1.5 M_{\odot}$ and $Z=0.01$, and are computed at three different initial velocities.
       The data points in the right-hand side panel are contained in the data set of the left-hand side panel.
}
     \label{magvsrot_2cols}
    \end{figure*}
    %
  
  \section{Models and comparison with data}
  \label{results}
  \subsection{Core rotation rate}
  In our models, two physical scenarios are considered (see Sect. \ref{methods}): transport only by hydrodynamical processes, and another one in which we consider also internal magnetic fields.
  In the following figures and discussion we refer to them as `rotation-only' or 'rotation+magnetic fields', respectively.
  In Fig. \ref{omegac_pi0_2cols}, we show the core rotation rate as a function of the buoyancy radius for a subset of our models with initial masses of $M=1.5, 1.7,$ and 1.9 $M_{\odot}$ and initial rotation rates of $\Omega_{\rm core}/2\pi=10, 20,$ and 30 $\mu$Hz, compared with the data available for our sample of $\gamma$ Dor stars.
  In these figures, we only include stars for which the uncertainty of the buoyancy radius is less than $\sigma_{P_{0}}=500$ s; models are computed from the ZAMS and the evolution goes from left to right.
  Both types of models have a decreasing core rotation rate during a large part of the MS, as evidenced by the large range in buoyancy radius during which the core spins down.
  However, internal magnetic fields lead to a faster spin down of the core right from the beginning of the evolution, erasing the initial spin-up seen in only-rotation models.
  Also, internal magnetic fields lead to a much lower core rotation rate towards the end of the MS, and a rather constant behaviour, while in rotation-only models, the core spins up by $\sim 50 \%$ and reaches much higher rates.
  Because of this, models with internal magnetic fields are in better agreement with the data since there are no fast rotators observed at low buoyancy radii.
  Also, the upper envelope of the distribution of core rotation rates is better reproduced when internal magnetic fields are included.
  And although we cover the mass-range expected for $\gamma$ Dor stars, the dependence with stellar mass in our models is weak.
  All this evidence supports the hypothesis of highly efficient AM transport during the MS.

  \subsection{Surface rotational velocity}
  \label{only_vsini}
  Since we have a good measurement of the surface rotational velocity for 104 stars in our sample of $\gamma$ Dor stars, we can also study the evolution of the surface velocity using the buoyancy radius as an evolutionary indicator; the comparison with our models is shown in Fig. \ref{vsini_pi0_2cols}.
  The surface velocity of our models is multiplied by a factor $\pi/4$ to average over the unknown inclination angle of the observed stars.
  The factor is equal to the mean value of $\langle \sin i \rangle$ where $i$ is the inclination angle, assuming an isotropic distribution of inclination angles whose probability distribution function is just $P(i)= 2\pi \sin i \ \rm{d}i$ \citep[][]{gray05}.
  
  The behaviour of the surface velocities when including internal magnetic fields is opposite to what is seen in Fig. \ref{omegac_pi0_2cols} for the rotation rate of the core.
  Compared to models with only hydrodynamical instabilities, internal magnetic fields lead to higher surface velocities, but the opposite occurs for the core (see Fig. \ref{omegac_pi0_2cols} \& \ref{vsini_pi0_2cols}); the core spins slower when internal magnetic fields are included.  
  This occurs because internal magnetic fields transport AM efficiently from the core to the surface, increasing the surface angular velocity, and hence (partially) counteracting the progressive slowdown due to the increase in radius during the MS.
    While rotation-only models have a decreasing surface velocity during most of the evolution, models with internal magnetic fields show a slight increase of the surface velocity at the beginning of their evolution, seen at high buoyancy radii.
    This mirrors the behaviour of the core when no internal magnetic fields are included (see left-hand side panel of Fig. \ref{omegac_pi0_2cols}).
    This is because, in models with internal magnetic fields, the AM transport is so efficient that the core and surface are coupled and hence rotate at nearly the same angular velocity through evolution.
    
    There is a lack of fast rotators at low buoyancy radii, which means that in our sample evolved stars are either not observed or do not reach high surface velocities.
    Still, it remains difficult to definitely conclude from Fig. \ref{vsini_pi0_2cols} that models with internal magnetic fields are clearly favoured over models with only hydrodynamical processes.

  \begin{figure}[ht!]
     \resizebox{\hsize}{!}{\includegraphics{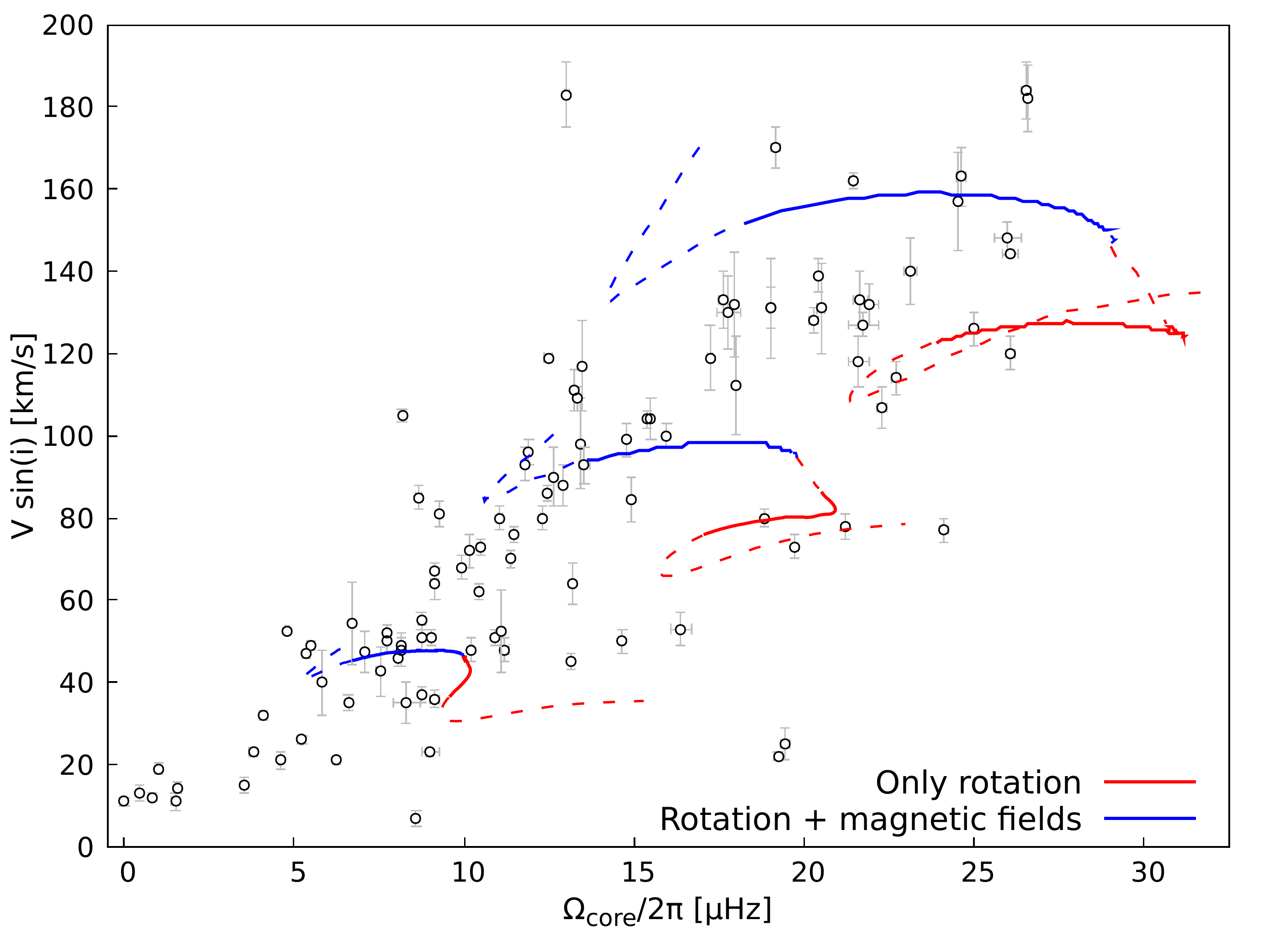}}
     \caption{Projected surface rotational velocity as a function of the core rotation rate.
       The lines correspond to models with an initial mass of $M=1.5 M_{\odot}$ and initial metallicity $Z=0.01$ shown only during the main sequence, computed with initial rotation rates of $\Omega_{\rm core}/2\pi = 10, 20$ and 30 $\mu$Hz, starting at the ZAMS.
       The blue and red lines indicate the type of AM transport assumed in each model: only hydrodynamical processes (red) or internal magnetic fields on top of hydrodynamical processes (blue).
       Data points are $\gamma$ Dor stars with constraints available from the literature (see Sect. \ref{data}).
       The solid lines represent the regions where the buoyancy radius of the models is in agreement with the mean values of the sample.}
     \label{vsini_omegac_1p5}
   \end{figure}
  %
 
    %
    \subsection{Combining surface- and core-information: spectroscopic rotational velocities}
    
    In the previous sections, we presented our individual study of the core rotation rate and  surface velocities as a function of the buoyancy radius in models with purely hydrodynamical processes and models with internal magnetic fields, and then compared their rotational properties to observational data.
    The effects of internal magnetic fields on the AM redistribution through the MS are illustrated in Fig. \ref{magvsrot_2cols} for a $1.5 M_{\odot}$ model computed with an initial metallicity of $Z=0.01$, which well represent the properties of the $\gamma$ Dor stars studied in this work (see Figs. \ref{histogram_feh}, \ref{histogram_mass}, and \ref{hr_diagram}).
    Now we use our sample of 104 stars for which we have constraints on surface rotational velocities, core rotation rates, and buoyancy radii to discern between our two types of models.
    To do this, we present the diagram of surface velocity as a function of the core rotation rate, containing the data available from our compilation and our models with only hydrodynamical transport of AM and models which also include internal magnetic fields.
    We show this diagram in Fig. \ref{vsini_omegac_1p5} for models with an initial mass of $M=1.5 M_{\odot}$ and three initial rotation rates of $\Omega_{\rm core}/2\pi = 10, 20$ and $30 \mu{\rm Hz}$.
    The models are shifted vertically by multiplying by $\pi/4$ the surface velocities to average over the unknown inclination angle of the stars.
    In this diagram, the two types of models start their evolution at the same point but right from the beginning they evolve in different directions; this is a consequence of the opposite behaviour in both core rotation rates and surface velocities obtained when internal magnetic fields are included (see Fig. \ref{magvsrot_2cols} and Sect. \ref{only_vsini}).
    We also use the values of the buoyancy radius to highlight the region of the evolutionary tracks which are in agreement with the mean values of the buoyancy radii of the sample of $\gamma$ Dor stars studied.
    This is shown by the solid lines, which show the part of the track where the buoyancy radius is $P_{0}=4092 \pm 575 s$, where the values correspond to the mean and the standard deviation of the sample, while the dashed lines correspond to buoyancy radii outside of that interval.
    
    Our models with purely hydrodynamical transport first evolve towards the right, increasing their core rotation rate until a turning point to then evolve towards the left, decreasing their core rotation rate.
    This behaviour is also seen in the left-hand side panel of Fig. \ref{omegac_pi0_2cols}.
    Their surface rotational velocity also decreases quite rapidly, which is the reason why for example for the models with an initial rotation rate of $\Omega_{\rm core}/2\pi \simeq 30 \mu{\rm Hz}$ the first part of the track does not satisfy the constraints on the buoyancy radii.
    As mentioned in Sect. \ref{only_vsini}, the surface velocity of rotation-only models decreases during most of the MS.
    It only increases towards the very end of the MS when the core hydrogen content is very low ($X_{\rm core} \lesssim 0.05$) and due to the low energy provided by the nuclear reactions in the core the whole star contracts, increasing both surface rotational velocities and core rotation rates.
    This is why the tracks move upwards and to the right when they approach the end of the MS, and it coincides with the hook seen in HR diagrams (see Fig. \ref{hr_diagram}).

    Our models with internal magnetic fields start their evolution in an opposite direction, both in core rotation rate and surface velocity.
    The models evolve to the left because the core rotation rate decreases almost monotonically until the star contracts due to the low central hydrogen content, close to the end of the MS.
    After this point, the core spins up imperatively since the whole star contracts, but since in this type of models core and surface are strongly coupled the spin-up is not so strong as in only-rotation models.
    As for the surface velocity, since magnetic models remain more coupled, their surface velocity does not change as much as in models without internal magnetic fields.
    As mentioned before, this occurs because the extraction of AM from the core to the surface counteracts the slowdown of the surface velocity due to the radius' expansion during the MS.
    Towards the end of the MS, just after the `hook' usually seen in the HR diagram, the surface contracts rapidly, leading to a sudden increase in the surface velocity, which is always present in our models irrespective of the efficiency of the internal AM transport.

  This roughly describes the reasons of the evolution in opposite directions in the diagram presented in Fig. \ref{vsini_omegac_1p5}.
  As for the comparison with data, we notice that most of the stars lie along a diagonal, indicating that the core rotation rate of the stars is well correlated with their surface rotational velocity.
  By comparing the behaviour of our only-rotation models we can discard an inefficient transport of AM since the models cannot reproduce the data for any initial rotation rate assumed. 
This occurs first because the core rotation rate increases at the beginning of the evolution, moving them away from the diagonal traced by the data.
But also because the surface velocity decreases during most of the evolution, causing the models to fall below the diagonal.
This illustrates the importance of combining different indicators which can trace also the structural evolution of the stars.
On the contrary, models with internal magnetic fields can reproduce the data because of the combined effect of a strong spin down of the core and little change in surface velocity, which result from the highly efficient AM transport by internal magnetic fields.

    We note however that this kind of diagram was already presented by \citet[][see their Fig. 7]{vanreeth18} but was not discussed in evolutionary terms.
    The behaviour seen for the data in Fig. \ref{vsini_omegac_1p5} is striking at first because we have to take into account that the inclination angle poses an uncertainty and should thus produce some random scatter along the natural relationship between both quantities.
    And although some stars fall below the diagonal, which could be explained by low inclination angles, the scatter remains weak.
    This could be also a bias toward high inclination angles as a product of selecting stars that show measurable rotational velocities by means of spectral lines broadening, and detectable g-modes in \textit{Kepler} light-curves. 
    According to \citet{li20} the detected modes are mostly sectoral, supporting the fact that probably the scatter in inclination angles is lower than expected from an isotropic distribution.
    However even considering that all stars are seen exactly equator on, models with inefficient transport of angular momentum would be in disagreement with data.

    This shows that a highly efficient AM transport is at work during the MS evolution of these stars, such that little to no differential rotation in radiative regions can develop through the whole MS.
    As will be discussed in Sect. \ref{uncertain}, such a conclusion is obtained independently from the different uncertainties in the input physics used in stellar models.

    \subsection{Combining surface- and core-information: surface rotation rates}
    \label{sect_surfrates}
    Another way to investigate the AM redistribution can be done by combining  projected surface rotational velocities from spectroscopy ($V \sin i$), effective temperatures ($T_{\rm eff}$) and bolometric luminosities ($L$).
    These parameters are available and were obtained with good accuracy for the sample of stars presented in Sect. \ref{data}.
    By combining those parameters, we can estimate the surface rotation rate of the stars as
    \begin{equation}
      \Omega_{\rm surf} \sin i \simeq \frac{(V \sin i)}{R} = \frac{(V \sin i)}{\sqrt{L}} T_{\rm eff}^2 \sqrt{4\pi\sigma_{\rm SB}}
      \label{eq_omsurfbb}
    \end{equation}
    where $\Omega_{\rm surf}$ is the surface rotation rate of the star and the rest of the symbols have their usual meaning.
    The stellar radii are then estimated using the luminosities provided by \citet{murphy19}.
    Since accuracy is important in this case, we choose the effective temperatures provided by the individual works mentioned in Sect. \ref{data}, which were all obtained from medium- to high-resolution spectra.    
    This is because the effective temperature of several of the stars provided by \citet{mathur17} were originally estimated with photometric magnitudes and served as input values for the \textit{Kepler} input catalog \citep{brown11}; in some other cases the effective temperature present in that catalogue was taken from results coming from low-resolution spectra \citep[e.g. LAMOST-DR1][]{luo15}.
    In addition to these choices to compute the stellar radius of the stars in our sample, when available, we use the radius obtained by \citet{mombarg21} for 37 of our stars.
    Hence, we argue that the data presented in this section are accurate and representative of the typical properties of $\gamma$ Dor stars.
    We note that for most of the stars presented in this section, the effective temperatures are provided by \citet{gebruers21}, which were obtained with high-resolution spectra and comprise a homogeneous sample.
    \begin{figure}[h!]
     \resizebox{\hsize}{!}{\includegraphics{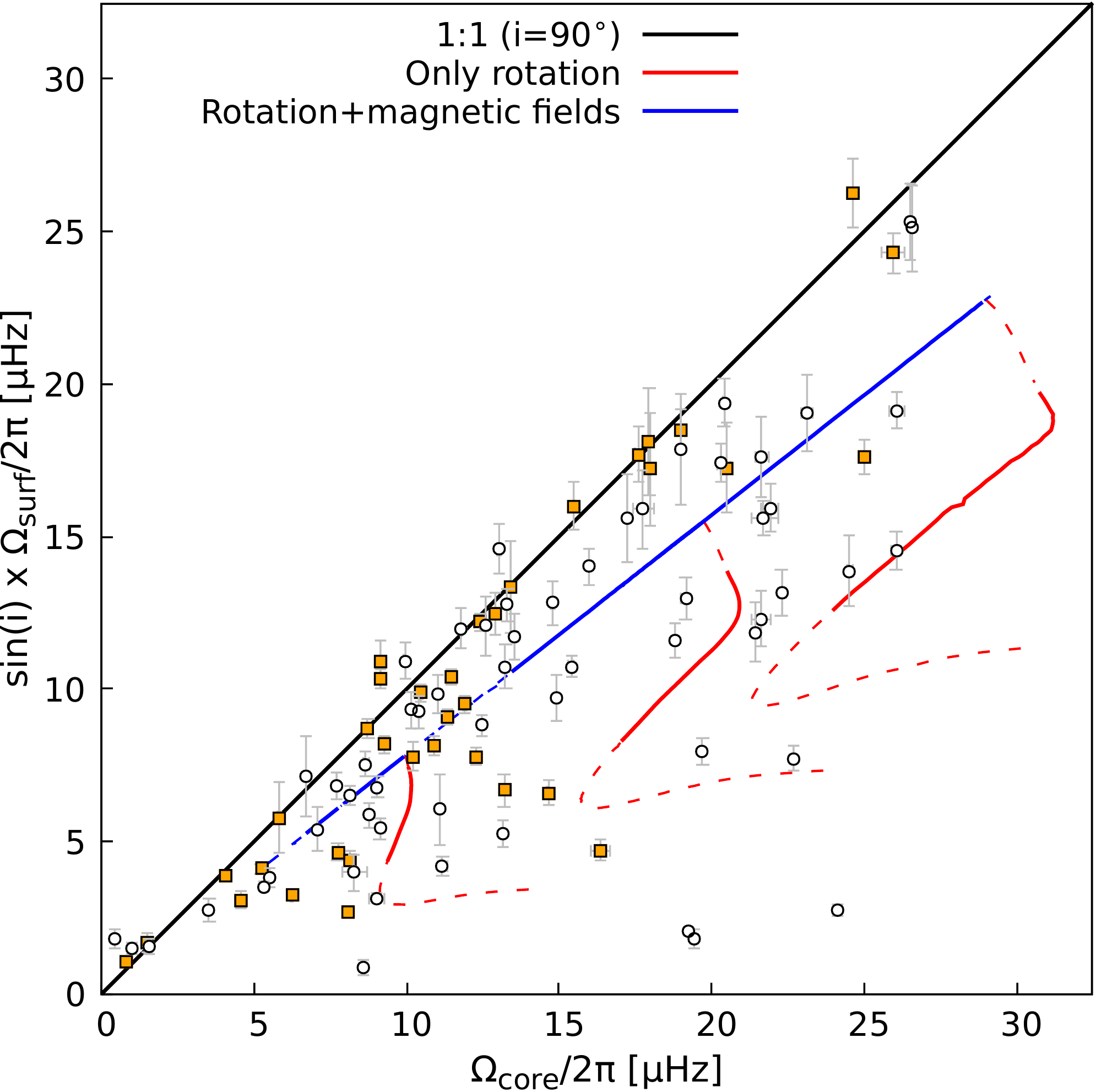}}
     \caption{Surface rotation rates as a function of the core rotation rate.
       The surface rotation rates are obtained by dividing the projected rotational surface velocities by the radius of each star.
       The dots represent the data set used in this work (see Sect. \ref{data}) while the red and blue lines have the same meaning as in Fig. \ref{vsini_omegac_1p5}, and the black line shows the one-to-one relation.
     }
     \label{omegas_omegac}
   \end{figure}
    This way we can compare surface- with core-rotation rates for the stars in our sample, which is shown in Fig. \ref{omegas_omegac} along with our stellar models.
    The stars whose radius was derived by \citet{mombarg21} are shown by orange squares, and for the rest of the stars represented by circles the radius was computed with the usual blackbody relation using Eq. \ref{eq_omsurfbb}.
    The inclination angle to obtain the exact surface rotation rate remains unknown.
    The error bars are estimated as the standard deviation with a standard propagation of errors neglecting correlations and taking into account the uncertainties in $T_{\rm eff}, V \sin i$, and $L$.
    The data are distributed in a similar way as when comparing core rotation rates with surface rotational velocities; a fast-spinning core is usually related to a fast-spinning surface.
    Some stars show low surface-rotation rates with rather high core-rotation rates which could be due to low inclination angles, and almost no stars appear to have a surface that rotates faster than its core.
    This is shown with help of the one-to-one line which would correspond to the location of solid-body rotating stars seen under inclination angles of $i=90 ^{\circ}$.
    Stars above the black line in Fig. \ref{omegas_omegac} should have a surface rotating faster than its core, irrespective of its inclination angle.
    Below this line, stars could either have a core that rotates faster than its surface, with a maximum inclination angle of $i=90^{\circ}$, or could have a surface rotating faster than its core with an inclination angle depending on the rotation contrast between surface and core; for example for a contrast of $\Omega_{\rm surf}/\Omega_{\rm core} \simeq 1.5$ an inclination angle of $i \lesssim 42^{\circ}$ would suffice to move the points below the one-to-one line.
    There are 11 stars which seem to have a surface that rotates faster than its core, seen at both low and high rotation rates.
    Overall most of the stars ($\sim 90 \%$) are found below the one-to-one line in Fig.~\ref{omegas_omegac}, indicating that it is unlikely that in our sample of stars the surface spins faster than their core.

    In Fig. \ref{omegas_omegac} we show the same stellar models as presented in Fig.~\ref{vsini_omegac_1p5}.
    The meaning of the different colours and type of lines is the same.
    The surface rotation rate of the models is also multiplied by a factor $\pi/4$ to average over the unknown inclination angles of the stars.
    Our models with purely hydrodynamical transport (red lines) develop differential rotation during the evolution with a core that rotates faster than its surface, this is the reason why the tracks extend both horizontal and vertically in this diagram.
    On the other hand, models with internal magnetic fields lie along a diagonal line because the angular velocity in these models is almost uniform, and so they lie along a line of rigid rotation whose slope is determined by the inclination angle assumed.
    More specifically, what we compare here is the mean rotation rate of the g-mode cavity with the rotation rate at the stellar surface.
    This means that in these models the angular velocity in the radiative interior is nearly constant.
    In particular, models with internal magnetic fields would always lie along a diagonal irrespective of their masses because in this figure we compare the rotation rates, so the change in radius due to different masses or metallicities is not seen.

    Taking into account this information provided by the models, and since most of the stars lie along a narrow diagonal band, we can interpret this as very efficient AM transport during the evolution.   
    The spread would correspond to different inclination angles, while the diagonal line traced by our model with internal magnetic fields would correspond to an inclination of $i \simeq 52^{\circ}$.
    Otherwise, if models developed strong differential rotation during the evolution, we would expect to see more stars on the lower-right part of this diagram, as not only the effect of the inclination angle would contribute to populate it but also the evolution of stars with different properties (such as masses, metallicities, initial velocities) would contribute to populate it.
    However, this region of the diagram is very weakly populated, which we interpret as evidence against models that develop strong differential rotation during the evolution, and hence against our models with purely hydrodynamical transport.
    In addition to this, we have to recall that all the stars studied here are g-mode pulsators and the visibility of the modes is not the same under different inclination angles.
    Nearly all stars presented in Figs. \ref{vsini_omegac_1p5} \& \ref{omegas_omegac} are pulsating in sectoral modes, and as pointed out by \citet{li20}, dipole prograde modes ($\ell =1, m=1$) have the largest observed amplitudes.
    This would support the fact that the stars are rather seen at high inclination angles since those modes have larger amplitudes at high inclination angles \citep{dziembowski07}.
    This provides further support to an efficient transport process of AM in stars during the MS.
    
  \section{Uncertainties in stellar modelling}
  \label{uncertain}
  Although our study supports the idea of a physical process able to transport AM very efficiently during the MS, we are aware that many uncertainties can impact the properties of stellar models.
  And even if they are not related to the transport of AM itself, due to the nature of our interdisciplinary study, it could lead to incorrect conclusions.
  We summarise here the possible uncertainties and their impact on our conclusions.

  \subsection{Overshooting}
\begin{figure}[htb!]
     \resizebox{\hsize}{!}{\includegraphics{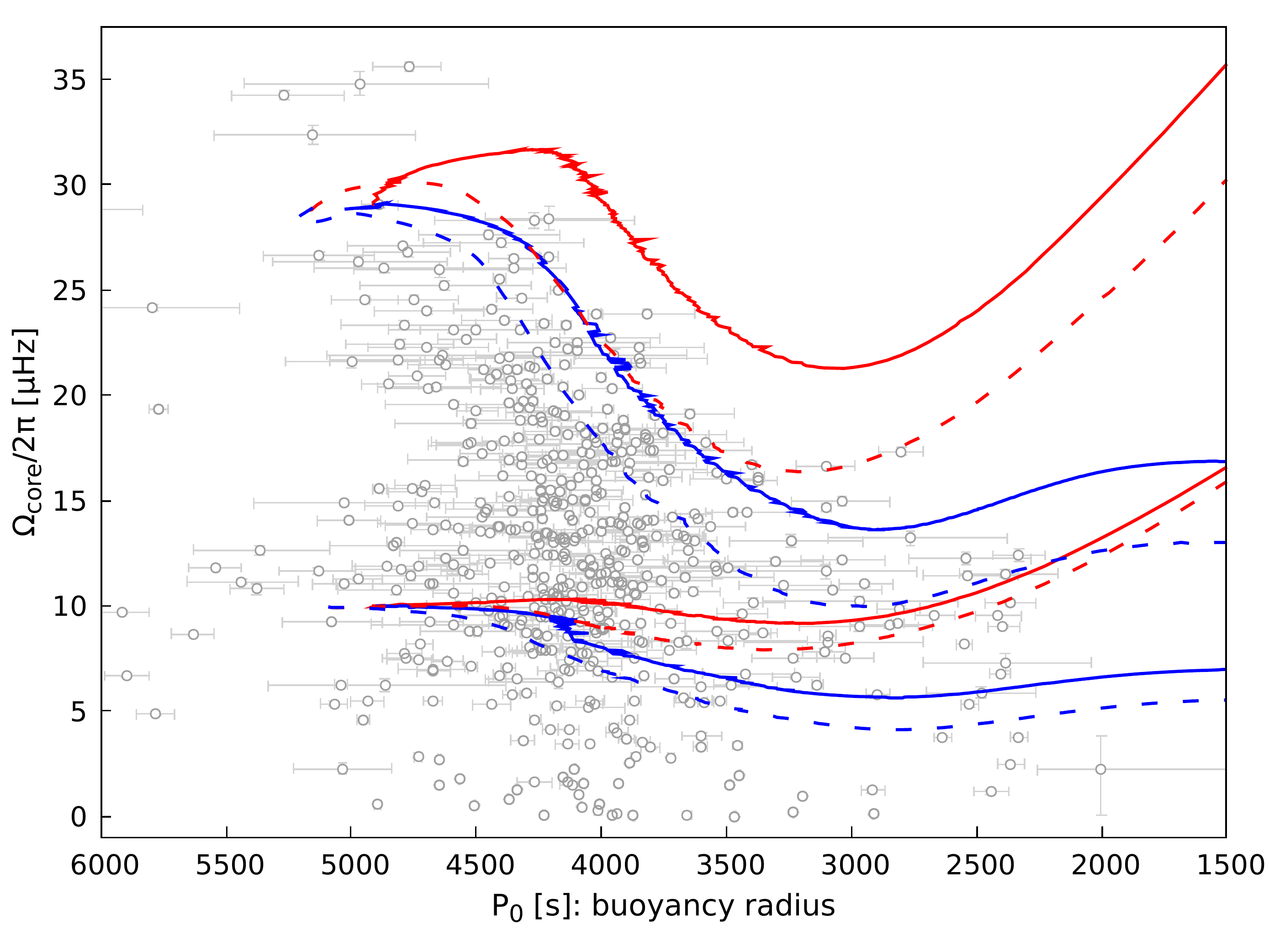}}
     \caption{Same as Fig. \ref{omegac_pi0_2cols} but showing the effect of the overshooting for models with transport only by hydrodynamical processes (red) and models that also include internal magnetic fields (blue).
       Solid lines correspond to models computed with an overshooting length of 0.05 $H_{\rm p}$ while dashed lines have an overshooting length of 0.2 $H_{\rm p}$.
       Models have an initial mass of  $M=1.5 M_{\odot}$.
       For clarity we only show the models computed at two different initial rotation rates.
     }
     \label{omegac_pi0_ov}
   \end{figure}

  One fundamental quantity in the modelling of MS low-mass stars with $M \gtrsim 1.2 M_{\odot}$ is the size of their convective core.
  It determines the luminosity of the star and hence its radius; a larger convective core leads to a more luminous star.
  Still, it is an uncertain quantity and many studies point out to a discrepancy between data and models computed with the standard Schwarzschild criterion of convection \citep[see e.g.][]{claret16}.
  To overcome this discrepancy stellar modellers usually extend the size of the convective core to account for an unknown mixing process between the convective core and radiative regions above it.
  In our models we extend the size of the convective core by $0.05 H_{\rm p}$ in our $1.5 M_{\odot}$ models and $0.1 H_{\rm p}$ for $M > 1.7 M_{\odot}$ adopting the step overshooting formalism.
  This proved to be a good fit for the observed width of open clusters in previous grids of models \citep[e.g.][]{ekstrom12}.
  To explore the possible impact of a larger convective core, we recomputed models with an increased overshooting strength of $0.2 H_{\rm p}$, which is quite a large value for the range of stellar masses considered in this work.
  Its effect on the evolution of the core rotation rate as a function of the buoyancy radius is shown in Fig.~\ref{omegac_pi0_ov} for our models with only hydrodynamical processes (red line) and models that also include internal magnetic fields (blue line).
  A larger overshooting strength leads to a higher buoyancy radius in both types of models and lower core rotation rates during the whole MS.
  However the evolution of the core rotation rate is not strongly affected.
  Models with internal magnetic fields still lead to lower core rotation rates towards the end of the MS while models with purely hydrodynamical transport lead to high core rotation rates at low buoyancy radii.
  Although a stronger overshooting could help reducing the core rotation rate of models without internal magnetic fields, it is not supported by the lack of fast high core rotation rates at low buoyancy radii.
  On the other hand, models with internal magnetic fields have a similar behaviour overall and even a strong increase of the overshooting does not lead to a disagreement with the data.  
    %

\begin{figure}[htb!]
     \resizebox{\hsize}{!}{\includegraphics{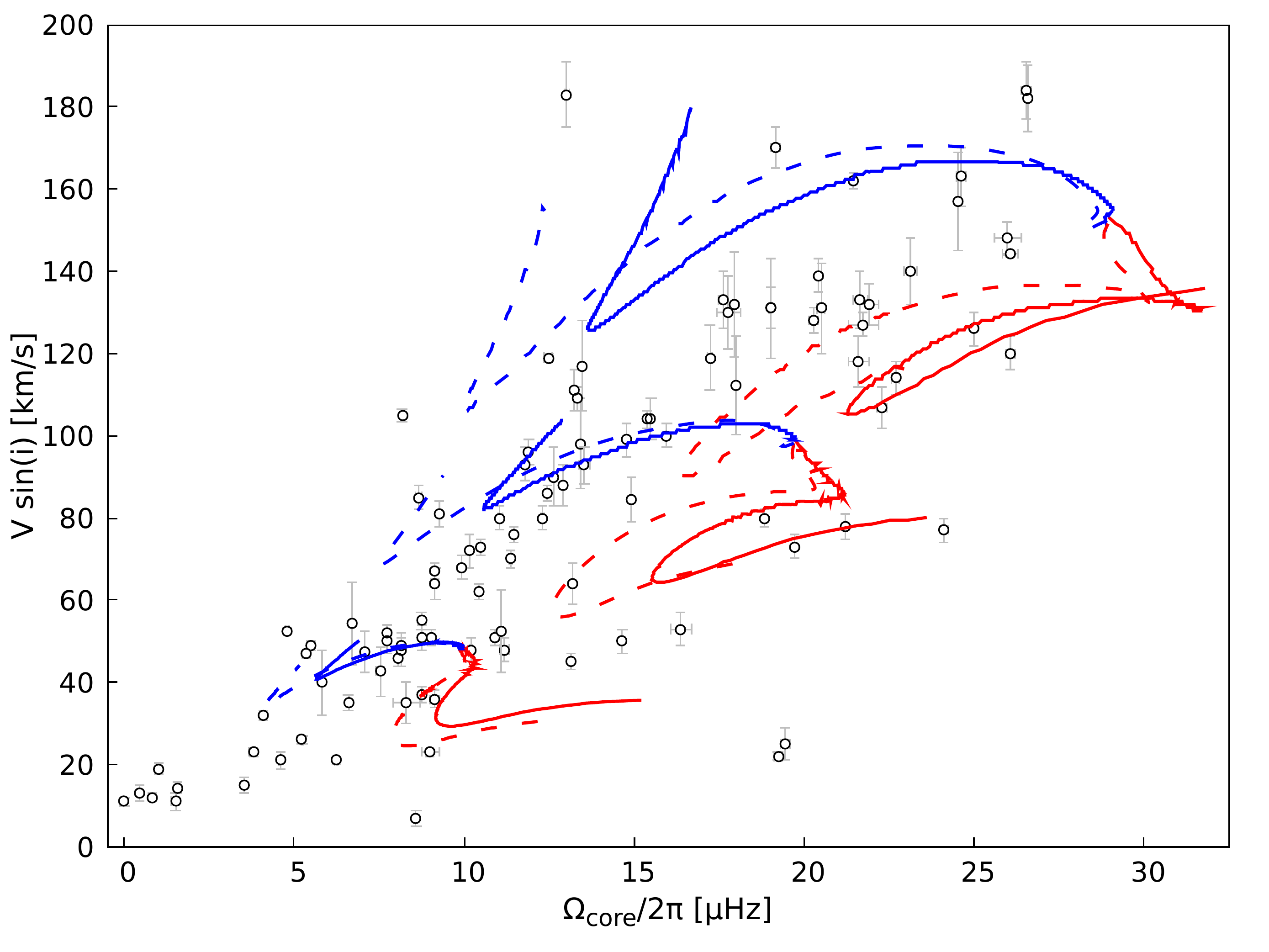}}
     \caption{Same as Fig. \ref{vsini_omegac_1p5} but showing the effect of the overshooting for models with transport only by hydrodynamical processes (red) and models that also include internal magnetic fields (blue).
       Solid lines correspond to models computed with an overshooting length of 0.05 $H_{\rm p}$ while dashed lines have an overshooting length of 0.2 $H_{\rm p}$.
       Models have an initial mass of  $M=1.5 M_{\odot}$.}
     \label{vsini_omegac_ov}
   \end{figure}

    We can also verify whether a more efficient overshooting can change our conclusions by comparing simultaneously the surface velocities and core rotation rates with our models.
    We show this comparison in Fig. \ref{vsini_omegac_ov} in the same way as in Fig.~\ref{omegac_pi0_ov} but including also the model with an initial rotation rate of $\Omega_{\rm core}/2\pi = 20 \mu {\rm Hz}$.
    The models with a more efficient overshooting (dashed lines) can achieve lower surface velocities during the MS because a larger convective core leads to higher luminosities and hence larger radii.
    The models with a stronger overshooting can approach the diagonal traced by the data due to their lower core rotation rates (see Fig. \ref{omegac_pi0_ov}).
    Still, it is not enough for models without internal magnetic fields to reproduce the data.
    Thus a different choice of the overshooting can not counteract the lack of AM transport in models with meridional circulation and shear instabilities during the MS.    
    %
    
    \subsection{Initial metallicity}
\begin{figure}[htb!]
     \resizebox{\hsize}{!}{\includegraphics{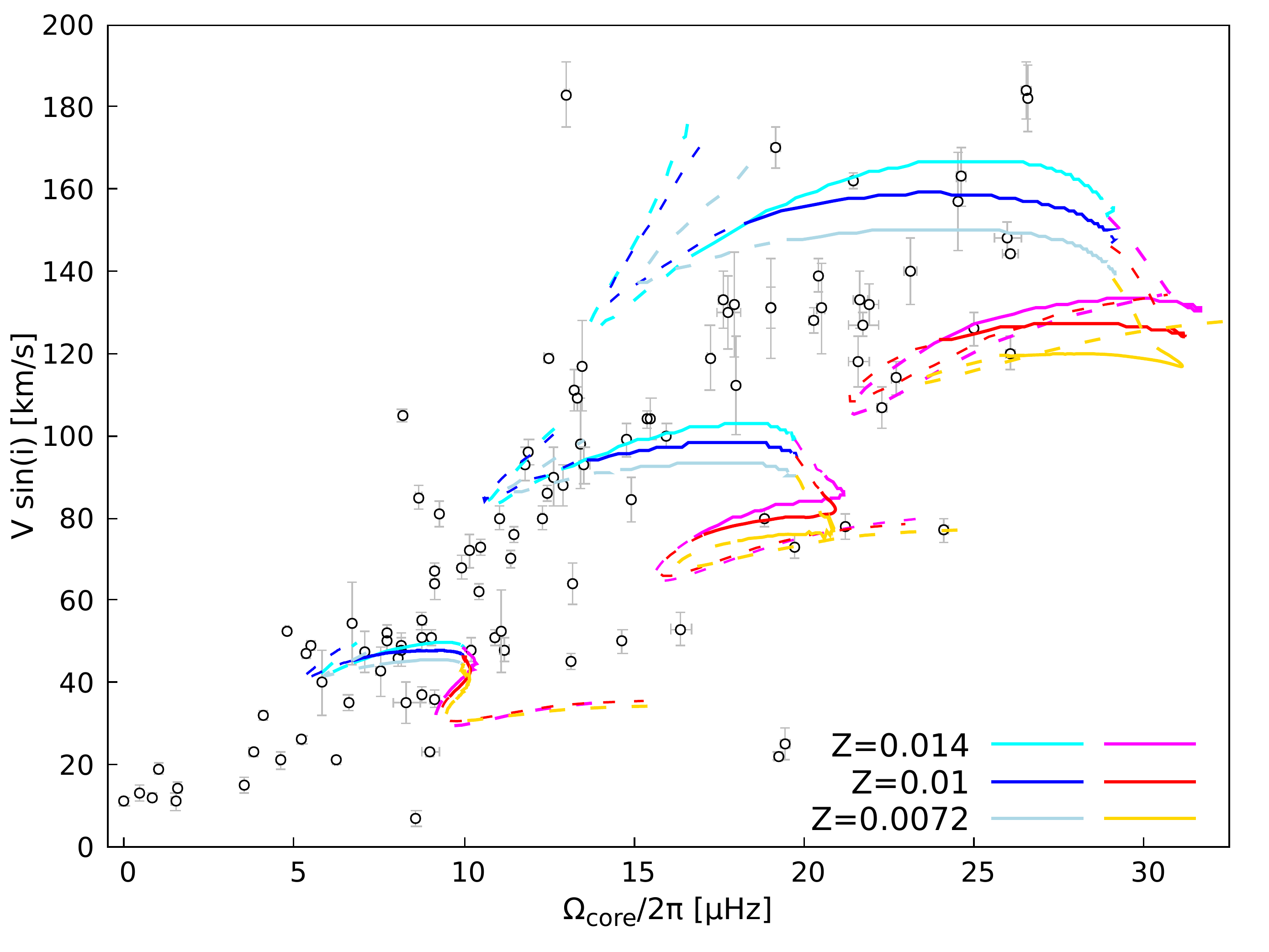}}
     \caption{Same as Fig. \ref{vsini_omegac_1p5} but showing the effect of different initial metallicities.
       Cyan, blue, and light-blue lines correspond to models computed with internal magnetic fields, whereas magenta, red, and yellow lines correspond to models computed only with hydrodynamical processes, hence without internal magnetic fields.
       Models are computed with initial rotation rates of $\Omega_{\rm core}/2\pi = 10, 20$ and $30 \mu$Hz.
       The dashed and solid parts of the line have the same meaning as in Fig. \ref{vsini_omegac_1p5}.
     }
     \label{vsini_omegac_met}
   \end{figure}

    In Fig. \ref{histogram_feh} we show the  metallicity distribution of our sample, which we use to choose the initial metallicity of our main models as $Z=0.01$.
    As discussed in Sect. \ref{data} the distribution spans a range of possible values, leading to a small uncertainty.
    This does not change significantly the behaviour of our evolutionary tracks nor the exact values of neither the core rotation rate nor the surface velocities.
    In Fig. \ref{vsini_omegac_met} we show models in the $V \sin i - \Omega_{\rm core}$ diagram starting at initial metallicities of $Z=0.0072, 0.01,$ and $0.014$ which correspond to [Fe/H]$ \simeq -0.3 \textrm{ to } 0$.
        All models have an initial mass of $M=1.5 M_{\odot}$ and initial rotation rates of $\Omega_{\rm core}/2\pi = 10 ,20,$ and $30 \mu$Hz.
    Models with higher metallicities in general have higher initial surface velocities for a fixed initial rotation rate because their radius is larger at the ZAMS.
    This is why the models seem to be vertically shifted with respect to each other.
    This shows that the effect of the metallicity is limited (irrespective of wheter internal magnetic fields are included) and cannot change the conclusions.
   
    \subsection{Initial distribution of AM}
    Since we compute all of our models from the ZAMS we must make some assumptions about the initial distribution of internal AM.
    In our models, we assume uniform rotation as an initial condition, i.e. solid-body rotation at the ZAMS.
    Usually after a few Myr the rotation profile reaches an equilibrium state as a consequence of the advective currents and the diffusive turbulent transport by the shear instability, and so the initial conditions do not have a significant impact on the structural evolution; thus for evolutionary purposes it is not a bad approximation.
    However in our study, the initial rotation profile is important because it defines the starting point in our $V \sin i - \Omega_{\rm core}$ diagram (e.g. Figs. \ref{vsini_omegac_1p5} \& \ref{vsini_omegac_ov}).
    An initial rotation profile where the core rotates faster than the surface would shift the starting point downwards and to the right in Fig. \ref{vsini_omegac_1p5}.
    On the other hand an initial rotation profile where the surface rotates faster than the core would shift it upwards and to the left.

    An initial rotation profile at ZAMS where the core rotates faster than the surface appears as the more plausible physical scenario \citep{haemmerle13}.
    Nonetheless the degree of differential rotation at the ZAMS can not attain large values, achieving at most a rotation contrast between convective core (cc) and surface of $\Omega_{\rm cc}/\Omega_{\rm surf} \simeq 1.5$ \citep{haemmerle13}.
    This initial choice would worsen the disagreement between data and models with purely hydrodynamical transport.
    This is because as explained above it would shift the starting point away from the diagonal traced by the data, and afterwards the evolution would be rather similar to a model starting with a uniform rotation profile.
    On the other hand, for models with internal magnetic fields the AM transport is so efficient that it returns practically instantaneously to its departure point of solid body rotation and would follow afterwards a rather normal evolution.
    This brings more evidence against models where only hydrodynamical processes transport AM in MS stars. 

\begin{figure}[htb!]
      \resizebox{\hsize}{!}{\includegraphics{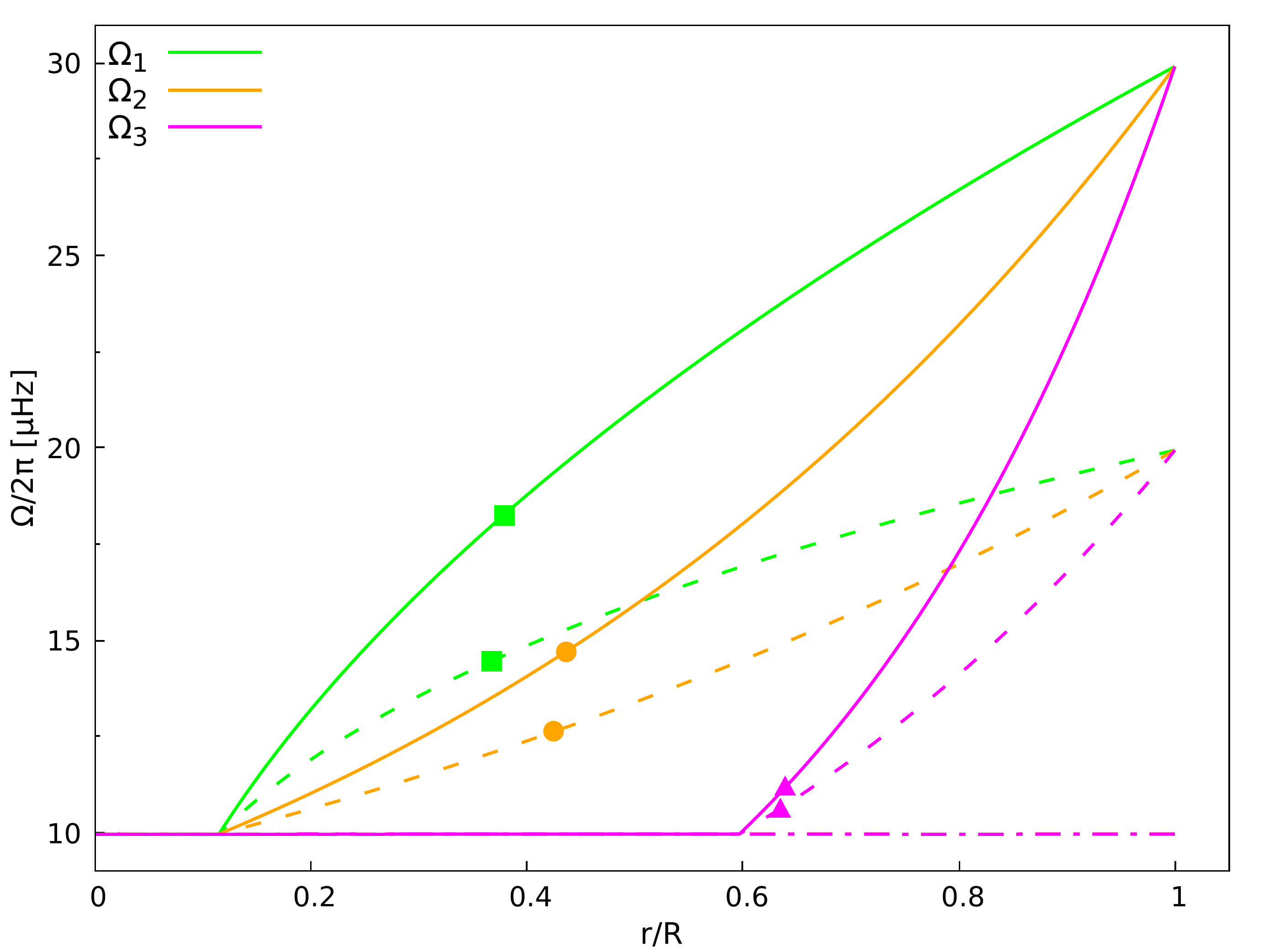}}
      \caption{Rotation profiles given by Eqs. \ref{omega_plaw}, \ref{omega_grow_shear}, and \ref{omega_solid_shear} at the ZAMS for a $1.5 M_{\odot}$ model for three different initial rotation ratios between convective core and surface of $\Omega_{\rm surf}/\Omega_{\rm cc} = 1, 2$, and 3, which correspond to the dot-dashed, dashed, and solid lines, respectively.
        The bullets shown on the rotation profiles show the location where the angular velocity is equal to the mean value of the g-mode cavity computed with Eq. \ref{eq_meanomega}.
     }
     \label{rotprofs_zams}
   \end{figure}

    If we now consider an initial rotation profile where the surface rotates faster than the core the starting point in Fig. \ref{vsini_omegac_1p5} could be shifted upwards and to the left.
    Under this assumption, models without internal magnetic fields could potentially reproduce the data.
    However such an initial rotation profile seems unlikely from a physical point of view, so we would expect stars to have rather shallow gradients at the ZAMS, which would then constrain the range of possible initial rotation contrasts in our models.
    To investigate this question, we constructed initial rotation profiles assuming different rotation laws: first employing a simple power law (Eq. \ref{omega_plaw})
    \begin{equation}
      \label{omega_plaw}
      \Omega_{1} (r)=K r^{q}
        \end{equation}
    a profile with increasing positive shear towards the surface (Eq. \ref{omega_grow_shear})
        \begin{equation}
      \label{omega_grow_shear}
      \Omega_{2} (r) = K_{1} r^{K_{2}r}
    \end{equation}
        and a profile where solid body rotation is assumed for $\simeq 60 \%$ of the star and afterwards the increasing shear law is used (Eq. \ref{omega_grow_shear})
        \begin{equation}
      \label{omega_solid_shear}
      \Omega_{3}(r)=
      \begin{cases}
        \Omega_{0}  & \text{if } r/R \le 0.6\\
        K_{1} r^{K_{2}r}  & \text{if } r/R  > 0.6.\\
      \end{cases}
    \end{equation}
            The variables $K, q, K_{1}, K_{2}$ and $\Omega_{0}$ are constants that are fixed by the initial angular velocities at the convective core and surface.
            The rotation profiles are shown in Fig. \ref{rotprofs_zams} for three different initial ratios at the ZAMS.
            These rotation profiles could for example result from accretion of matter in (approximately) Keplerian orbits around the star; in this case even a relatively low ( $\sim 3 - 5$ \% of the stellar mass) content of matter can spin up the surface of the star to critical velocities \citep{packet81}.
              These episodes of accretion could be for example the result of mass-transfer from a companion or accretion of a remnant disk.
              And depending on the efficiency of the AM redistribution in the near-surface layers, different slopes for the rotation profiles could be expected \citep[e.g.][]{staritsin21}.
            We note that in our framework this would be valid if the star has already a radiative envelope.

    Since the rotation rate probed by gravity modes is an averaged rotation rate outside of the convective core over the g-mode cavity, the shape of the rotation profile can change the near-core rotation rate computed with Eq. \ref{eq_meanomega}.
    This is illustrated in Fig. \ref{rotprofs_zams} by bullets over the rotation profiles where we show the location of the star where the angular velocity would equal the mean angular velocity of the g-mode cavity for each profile.
      For the flat profile no bullet is shown simply because the mean angular velocity is the same as in the whole star.
      This shows how the mean core rotation rate probed by g-modes highly depends on the shape of the profile (but not the region probed) at the ZAMS.
    This is why it is not enough to consider only a different rotation contrast between the center of the star (i.e. the convective core) and surface at the ZAMS.
    To verify if it is possible for our rotation-only models to start their evolution from the left part of the diagonal traced by the data (see Fig. \ref{vsini_omegac_1p5}) we recomputed both the mean core rotation rate with Eq. \ref{eq_meanomega} and the surface rotational velocity using the structure of the models at the ZAMS for the different rotation profiles assumed.
    
\begin{figure}[htb!]
      \resizebox{\hsize}{!}{\includegraphics{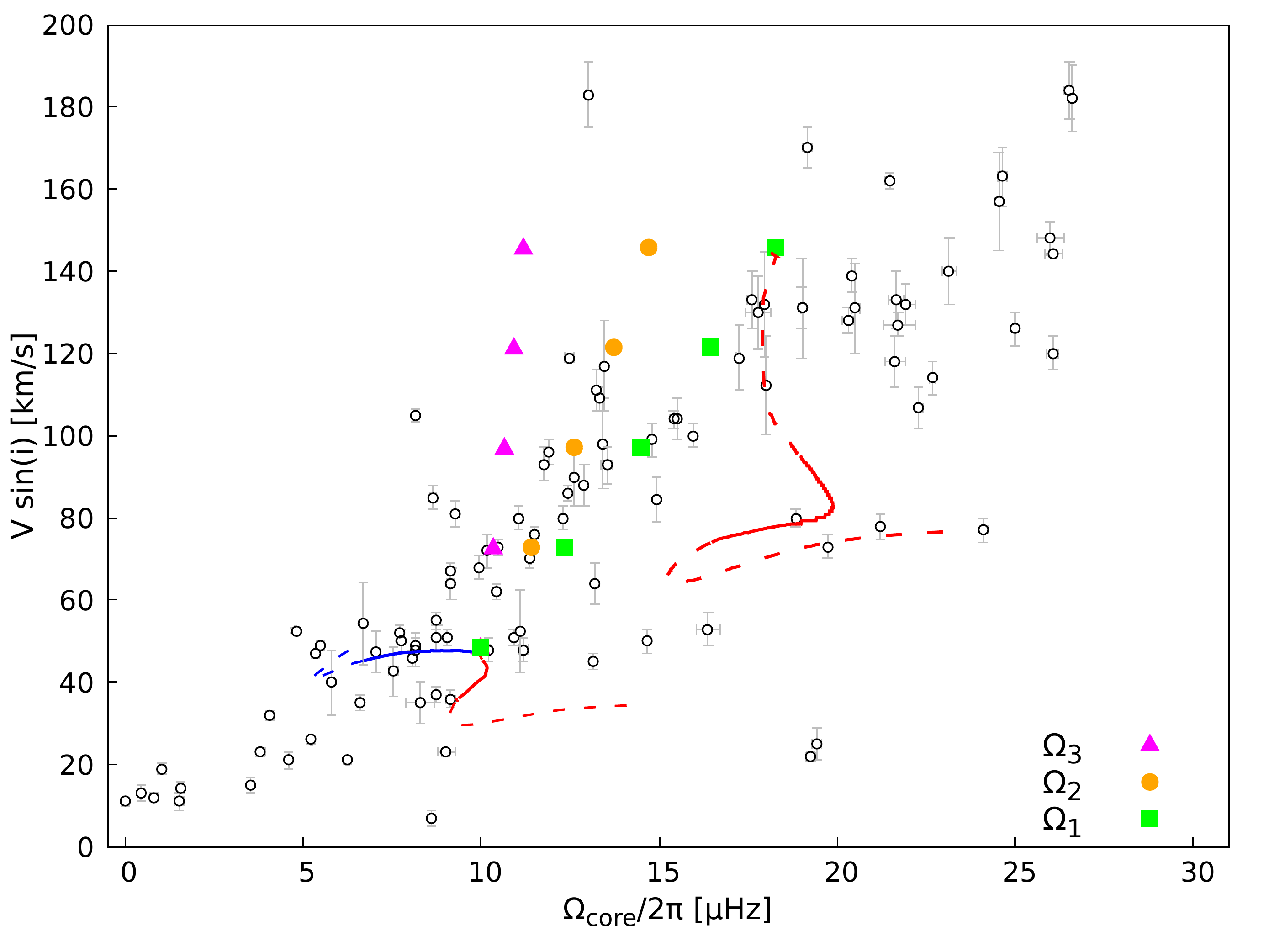}}
      \caption{Same as Fig. \ref{vsini_omegac_1p5} but showing the effect of assuming different initial rotational profiles.
        The points correspond to the ZAMS location of models starting with an initial rotation rate of $\Omega_{\rm cc}/2\pi= 10 \mu$Hz but assuming different degrees of differential rotation and rotation profiles (see Fig. \ref{rotprofs_zams}).
        The points are shown for initial contrasts of $\Omega_{\rm surf}/\Omega_{\rm cc}=1, 1.5, 2, 2.5$ and 3, where the solid-body rotation profile ($\Omega_{\rm surf}/\Omega_{\rm cc}=1$) coincides with the starting point of the models shown in blue and red lines.
        The red-line model starting from the upper green square was computed without internal magnetic fields and with an initial surface rotation rate three times higher than its core.
     }
     \label{vsini_omegac_diffzams}
   \end{figure}

    To achieve a starting point on the left of the diagonal traced by the data it is necessary to start with a low rotation rate in the convective core.
    If otherwise we assume a fast rotating convective core spinning at for example $\Omega_{\rm cc}/2\pi \simeq 30 \mu {\rm Hz}$, the surface velocities would be too high leading to a disagreement with the data.
    Then starting with a model with a convective core rotating at $\Omega_{\rm cc}/2\pi \simeq 10 \mu {\rm Hz}$ using different initial rotation profiles, we verify if it is possible to start the evolution on the left part of the diagonal traced by the data in Fig. \ref{vsini_omegac_1p5}.
    To achieve this, initial rotation contrasts of at least $\Omega_{\rm surf}/\Omega_{\rm cc} \gtrsim 3, 2.5$ and $1.8$ would be needed with the rotation laws $\Omega_{1}, \Omega_{2}$ and $\Omega_{3}$, respectively, with $\Omega_{\rm cc}$ the rotation rate of the convective core.
    This is illustrated in Fig. \ref{vsini_omegac_diffzams}, where we show the ZAMS point in the $V \sin i - \Omega_{\rm core}$ diagram for the different rotation profiles explored (see Eqs. \ref{omega_plaw}, \ref{omega_grow_shear}, and \ref{omega_solid_shear}), and initial rotation contrasts of $\Omega_{\rm surf}/\Omega_{\rm cc} = 1, 1.5, 2, 2.5,$ and 3.
    These points converge for a solid-body rotation profile, and thus are located at the ZAMS point of our models computed with an initial rotation rate of $\Omega_{\rm core}/2\pi= 10\mu$Hz.
    However, starting with a rotation profile where the surface rotates faster than the g-mode cavity, would lead to a lack of slow-rotators, both in near core-rotation rate and surface rotational velocity.
    This is not supported by the current data.

    Those ratios are rather large and it would remain difficult to invoke a physical scenario leading to these initial distributions, since accretion during the pre-MS would lead to a core spinning only slightly faster than its surface \citep{haemmerle13}.
    Moreover, once we compute the evolution of models where the initial rotation profile is modified, the evolutionary behaviour of both core rotation rate and surface velocity does not change; only-rotation models still have decreasing surface velocities and increasing core rotation rates, moving them to the lower-right in Figs. \ref{vsini_omegac_1p5} \& \ref{omegas_omegac}.
    While for models with internal magnetic fields, as mentioned before, the transport is so efficient that they quickly lead to uniform rotation and return to their normal evolution, just as if they were computed with a uniform rotation profile at the ZAMS.
    We also show in Fig. \ref{omegas_omegac} that the data do not support stars with a surface rotating faster than its core.   
    Taking all these arguments into account we deem unlikely that our models with purely hydrodynamical transport of AM can reproduce the data and so the conclusions remain robust regardless of the evolutionary history during the stellar formation and further pre-MS evolution.

    As an additional test, we check how a model with an initial strong degree of differential rotation at the ZAMS would evolve and whether it could reproduce the data.
      We do this by using the $\Omega_{1}$ profile given by Eq. \ref{omega_plaw} with an initial contrast of $\Omega_{\rm surf}/\Omega_{\rm cc} = 3$ (i.e. surface rotating faster than core) as an initial condition at the ZAMS, and then we compute the evolution with purely hydrodynamical AM transport.
      This model is shown in Fig. \ref{vsini_omegac_diffzams} starting from the green point above the diagonal traced by the data, which corresponds to the ZAMS location of a model star with such rotation contrast and rotation law.
      Although the model can simultaneously reproduce the surface rotational velocities and core rotation rates in the early MS, the buoyancy radius of the model in these parts is not in agreement with the mean observed values.
      And as shown by the solid lines, in this diagram the regions probed by our `only-rotation' models remains the same (i.e. below the diagonal traced by the data) and are thus in disagreement with the data.
      The fact that the buoyancy radius of this model is not in agreement with the data during the early MS occurs because the rotation profile readjusts rapidly during the evolution as a product mainly of the AM advection by meridional circulation, leading to a strong decrease in the surface rotational velocities.
      The evolution of the rotation profiles is shown in Fig. \ref{rotprof_r03}. 
    
    \section{Additional evidence and other indicators of evolution}
    \label{additional_evidence}
  We presented our main results in the previous sections using the buoyancy radius as a main indicator of the evolution during the MS.
  This is convenient since in the range of metallicities of our sample (see Fig. \ref{histogram_feh}) the buoyancy radius is weakly dependent on the metallicity (see Fig. 2 of \citet{ouazzani19}), and since we have an estimate of most likely stellar masses of our sample (see Fig. \ref{histogram_mass})  we can better distinguish between physical scenarios during stellar evolution.
  But we can also try to use other indicators and check how well our conclusions hold when confronted to independent indicators of evolution during the MS, which we discuss here.

  \subsection{Stellar clusters and asteroseismic inferences}
\begin{figure}[htb!]
     \resizebox{\hsize}{!}{\includegraphics{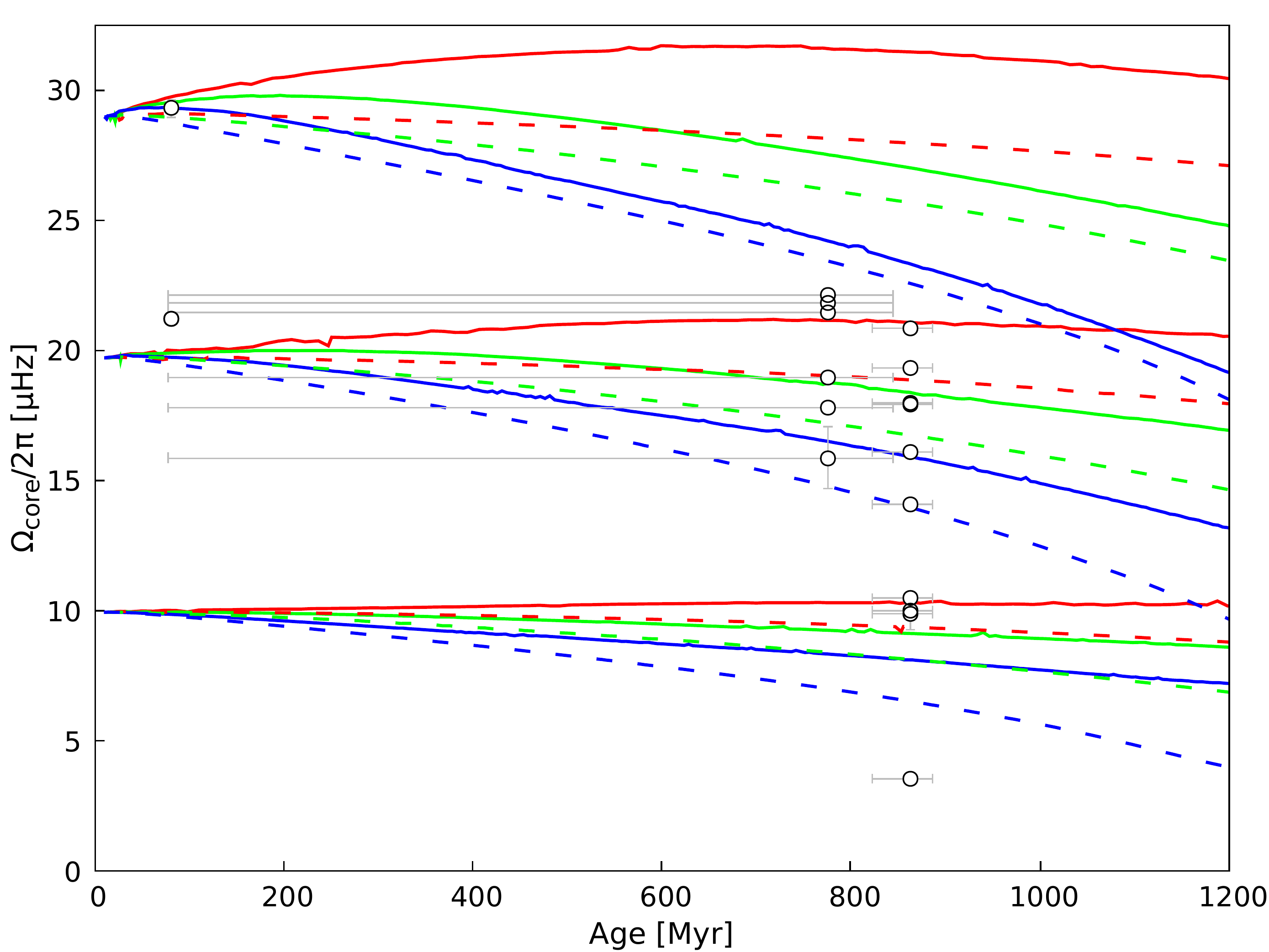}}
     \caption{Core rotation rate as a function of the age.
       The data points correspond to the $\gamma$ Dor stars in our sample that belong to the open clusters NGC6811, NGC6866, and ASCC108.
       The lines correspond to our models with only hydrodynamical processes (solid) and models including also internal magnetic fields (dashed).
       The different colours correspond to models with initial masses of $M=$1.5 (red), 1.7 (green), and 1.9 (blue) $M_{\odot}$; all computed at solar metallicity.
     }
     \label{omegac_age}
   \end{figure}

    We checked for membership of the stars in our sample to stellar clusters in the \textit{Kepler} field and we found 24 stars that belong to open clusters, namely NGC6811, NGC6866, and ASCC108.
    All three clusters are well characterised in terms of membership with aid of astrometric information from GAIA-DR2 \citep{cantat-gaudin18,cantat-gaudin20}, thus their ages can be reliably estimated with stellar isochrones.
    We use the ages derived by \citet{bossini19} for the three stellar clusters, and select only stars with a membership probability higher than 60 \%, based on the values given by \citet{cantat-gaudin20}.
  With this information we can give a reliable estimate of the age for 19 of our $\gamma$ Dor stars.
  And since they belong to three different open clusters we can infer the core rotation rate at three different ages for stars of different masses.
  The ages of the stellar clusters are  $t \simeq 81.5, 776,$ and $863$ Myr  for ASCC108, NGC6866,  and NGC6811, respectively, and have metallicities close to solar \citep{bossini19}.
  Then we can compare the core rotation rate of our models as a function of the absolute age, which is shown in Fig. \ref{omegac_age}.
      Although we see that older stars tend to have lower core rotation rates, it is still uncertain whether their cores should be slowing down during evolution and at which rate; based solely on Fig. \ref{omegac_age}.
      This is because there are very few stars with these kind of constraints available in the literature, and because the absolute age is not a good indicator to distinguish between models with either inefficient or efficient AM transport as evidenced by the mild difference in core rotation rate between both types of models in this figure.
      
\begin{figure}[htb!]
     \resizebox{\hsize}{!}{\includegraphics{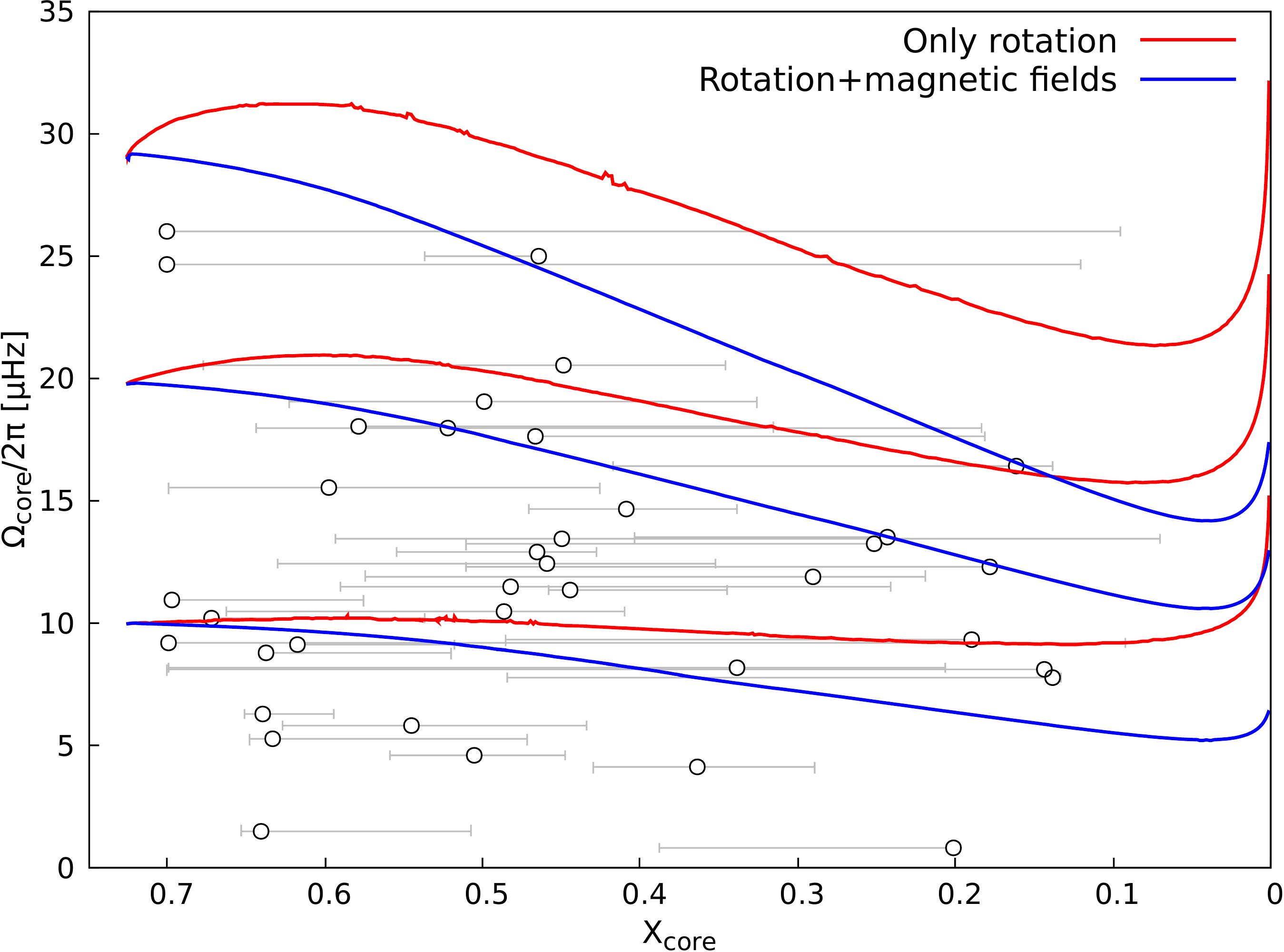}}
     \caption{Core rotation rate as a function of the mass fraction of hydrogen in the core.
       The data points correspond to $\gamma$ Dor stars with inferences on the central hydrogen content from \citet{mombarg21}.
       The lines show our models computed at three different initial rotation rates with only hydrodynamical processes (red) or including also internal magnetic fields (blue).
     }
     \label{omegac_xc}
   \end{figure}

      Besides, \citet{mombarg21} estimated the central hydrogen content of 37 stars present in our sample.
      We compare our two types of models against these constraints in Fig. \ref{omegac_xc}.
      In this figure we show our models computed with an initial mass of $M= 1.5 M_{\odot}$ and metallicity of $Z=0.01$, since the inferred mass of several of the stars in this sample lies around $M \simeq 1.5 M_{\odot}$ (see Table B.1 of \citet{mombarg21}) and the behaviour of the core rotation rate is weakly dependent on the metallicity for the stars in our sample.
      When comparing the core rotation rate as a function of the central hydrogen content there is an apparent progressive slow down towards the end of the MS, with a clear lack of fast rotators near the end of the MS seen in the data.
      This favors models with AM transport by magnetic instabilities but without clearly rejecting only-rotation models.
     %
      %

      \subsection{Evolved stars close to the end of the main sequence}
      Combining the different indicators presented, we finally selected stars that are most likely to be near the end of the MS.
      This would be decisive, since depending on the efficiency of the AM transport, stars are expected to evolve differently in both core rotation rate and surface velocity as shown in Fig. \ref{vsini_omegac_1p5}.
      In this figure, we would expect to find evolved stars mainly on the left part of the diagonal traced by the data if the AM transport was very efficient.
      On the other hand, if transport was inefficient during the evolution, they would have to lie mainly below the diagonal.
      This can also be verified using Fig. \ref{omegas_omegac}, but in this diagram, if AM transport was efficient, it could not be distinguished at first since all stars should line up along a diagonal depending on their inclination angles.
      Nonetheless, they should preferentially be grouped towards the lower left part of the diagram, since their core (and hence surface) should spin down through evolution as a result of their increase in size and efficient AM transport.
      
\begin{figure}[h!]
     \resizebox{\hsize}{!}{\includegraphics{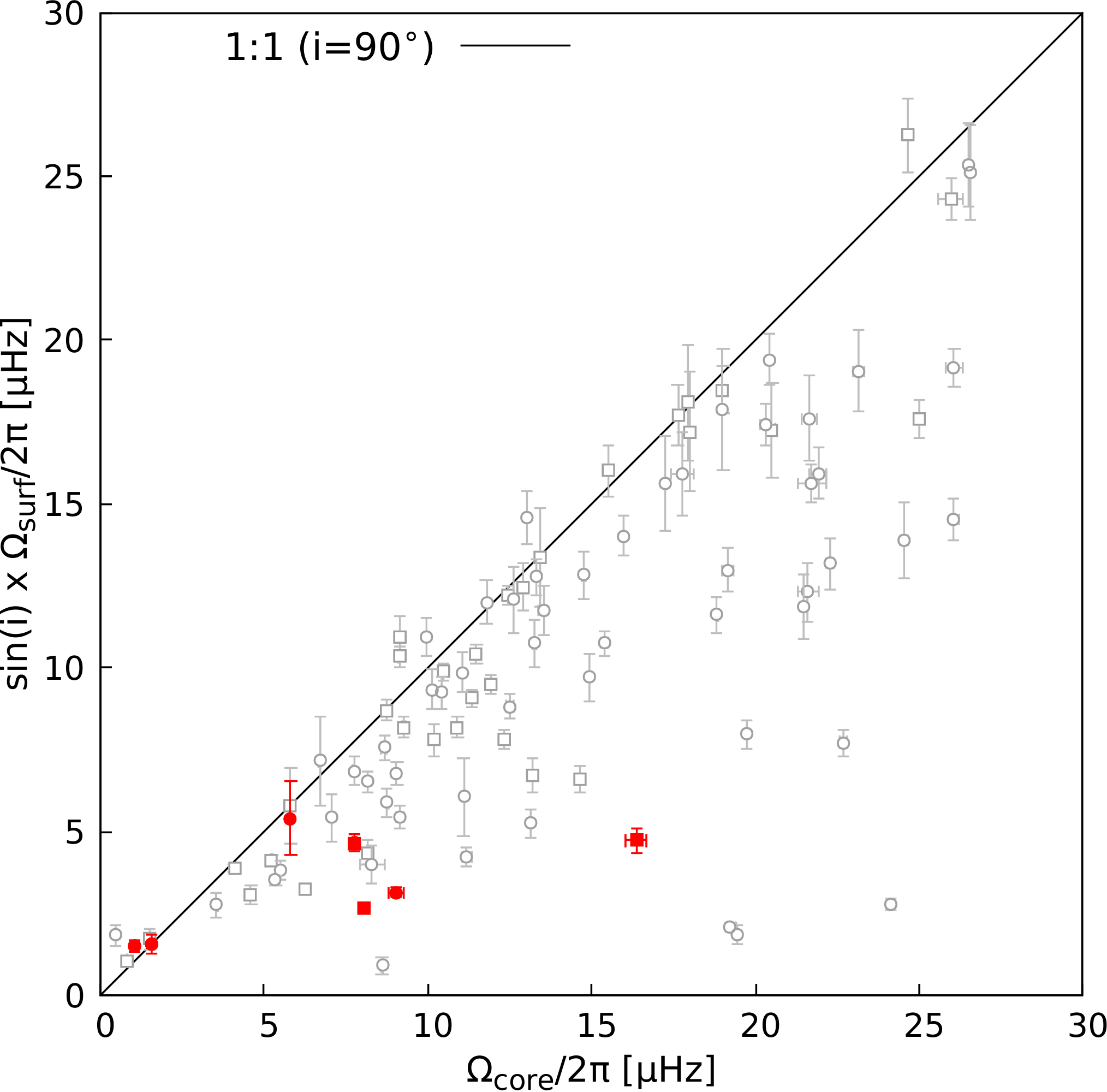}}
     \caption{Surface rotation rates as a function of the core rotation rate (see Sect. \ref{sect_surfrates}).
       The red symbols correspond to stars close to the end of the main sequence.
       The black solid line indicates the one-to-one line.
     }
     \label{evolved_stars}
   \end{figure}

      We only find 7 stars likely to be close to the end of the MS, based on their low core-hydrogen content \citep{schmid15,schmid16,mombarg21} and on the low values of their buoyancy radii.
      These stars are: KIC6678174, KIC7380501, KIC10467146, KIC11080103, KIC6519869, KIC10080943A, and KIC10080943B.
      We show their positions with respect to the whole sample in Fig. \ref{evolved_stars}.
      The first three stars mentioned have an inferred core hydrogen mass fraction lower than $X_{\rm core} \simeq 0.15$ \citep{mombarg21}.
      KIC11080103 and  KIC6519869 have buoyancy radii lower than $P_{0} = 3000 s$ and so they are expected to be close to the end of the MS.
      And the stars in the binary system KIC10080943 have both a low core-hydrogen content and low buoyancy radius \citep{schmid16}.

      Four of these stars (KIC7380501, KIC11080103, KIC10080943A, and KIC10080943B) are consistent with efficient AM transport because they lie close to the diagonal of solid-body rotation for stars seen at an inclination of $i=90^{\circ}$ in Fig. \ref{evolved_stars}.
      Although this mildly supports efficient transport during the MS, we argue that it strongly rejects inefficient transport during the evolution.
      If AM transport was inefficient, as in the case of our rotation-only models, we would expect all stars to lie in the lower part of Fig. \ref{evolved_stars}, irrespective of their inclination angles; a low inclination angle would only move them to lower parts of the diagram.
      This is because differential rotation should then develop, leading to a surface that rotates slower than the inner regions, and hence moving the position of the stars to the lower part of Fig. \ref{evolved_stars}.
      This can be seen for example comparing the expected position of the evolved stars close to the end of the MS in Fig. \ref{omegas_omegac} for our rotation-only models.
      Another possibility would be that actually the stars begun their evolution with strong differential rotation between core and surface, having a surface spinning faster than the core, and are then heading to the right part of the diagrams (i.e. developing a core that rotates faster than the surface).
      However, we disregard this scenario since due to meridional circulation, the rotation profile is expected to quickly converge to a profile where the core rotates faster than the surface (see Appendix \ref{app_diffzams} and Fig. \ref{vsini_omegac_diffzams}).
      For a starting rotation contrast of $\Omega_{\rm surf}/\Omega_{\rm core}=1.5$, our models achieve this once the hydrogen in the core drops to a mass-fraction of just $X_{\rm core} \simeq 0.65$.
      And since in general the data indicate that the stars do not have a surface spinning faster than their core (see bottom panel of Fig. \ref{evolved_stars}), we find it highly unlikely that near the end of the MS the stars could be in this configuration.
      We note however, that KIC10080943B seems to have a surface rotating faster than its core (see Fig. \ref{evolved_stars}).

      The other three stars (KIC6678174, KIC10467146, and KIC6519869) have low projected rotational velocities and are consistent with either low inclination angles and efficient AM transport, or, moderate to high inclination angles and inefficient AM transport.
      If they were rotating as solid bodies, then their inclination angles would be in the range $15^{\circ} \lesssim i \lesssim 35^{\circ}$.
      If we assume that the AM transport was inefficient during the evolution of these three stars and that the inclination angles are moderate to high ($i \gtrsim 52^{\circ}$), then the rotation contrast between the near-core regions and the surface should be in the range $\Omega_{\rm core}/\Omega_{\rm surf} \simeq 2.3 - 2.7$.
      This is in fragile agreement with our rotation-only models since the contrast achieved is at most $\Omega_{\rm core}/\Omega_{\rm surf} \simeq 2.0$ and $2.5$ when the central hydrogen mass-fraction reaches $X_{\rm core} \simeq 0.1$ for models with initial rotation rates of $\Omega_{\rm core}/2\pi = 10$ and $20 \mu{\rm Hz}$, respectively; and it is roughly independent of the initial mass.
      Although we are more inclined to say that these stars are not well explained by an evolutionary scenario where the AM transport is inefficient, we cannot entirely rule it out.

      None of the stars discussed above have reported rotational modulation in their light curves \citep{li20} and are not reported as eclipsing binaries \citep{kirk16}, although KIC10080943 is a double-lined spectroscopic binary system \citep{schmid16}.
      To give a more solid conclusion we would need to have more stars likely to be close to the end of the MS.
      But the four stars in agreement with efficient transport, if proved to be true evolved stars with rather high inclination angles, would not only give more support to an efficient transport in the MS but also to a process able to act during the whole MS.
      Because it would still be possible that the efficiency of the physical process decreases at some point during the evolution, but in that case they should quickly develop differential rotation and hence move to another location in the diagram presented, which is not supported by these stars.
    
      \subsection{Surface rotation rate from modulation of light-curves}
        For some $\gamma$ Dor stars there is evidence of modulation in the light curves that could potentially be a signature of magnetic spots at their surface and so provide direct measurements of their surface rotation rate.
      This would remove the uncertainty due to the unknown inclination angles present in the surface rotational velocities obtained with spectroscopy.
      \citet{li20} found signatures of surface modulation in the light curves of 58 out of the 611 stars that they analysed, and obtained their respective surface rotation rates concluding that they rotate nearly rigidly.
      Although this is a precise constraint on the nature of the AM transport in stellar interiors, different interpretations of the cause of the modulation in this group of stars blur the picture \citep[see][]{lee20,henriksen23}.
      Moreover, the detection of surface modulation in stars with a surface rotating slower than the near-core regions may have been overlooked due to selection effects \citep{li20} affecting so the conclusion on the efficiency of AM transport in stellar interiors; this does not occur if we rely on spectroscopic rotational velocities.
      
\begin{figure}[htb!]
        \resizebox{\hsize}{!}{\includegraphics{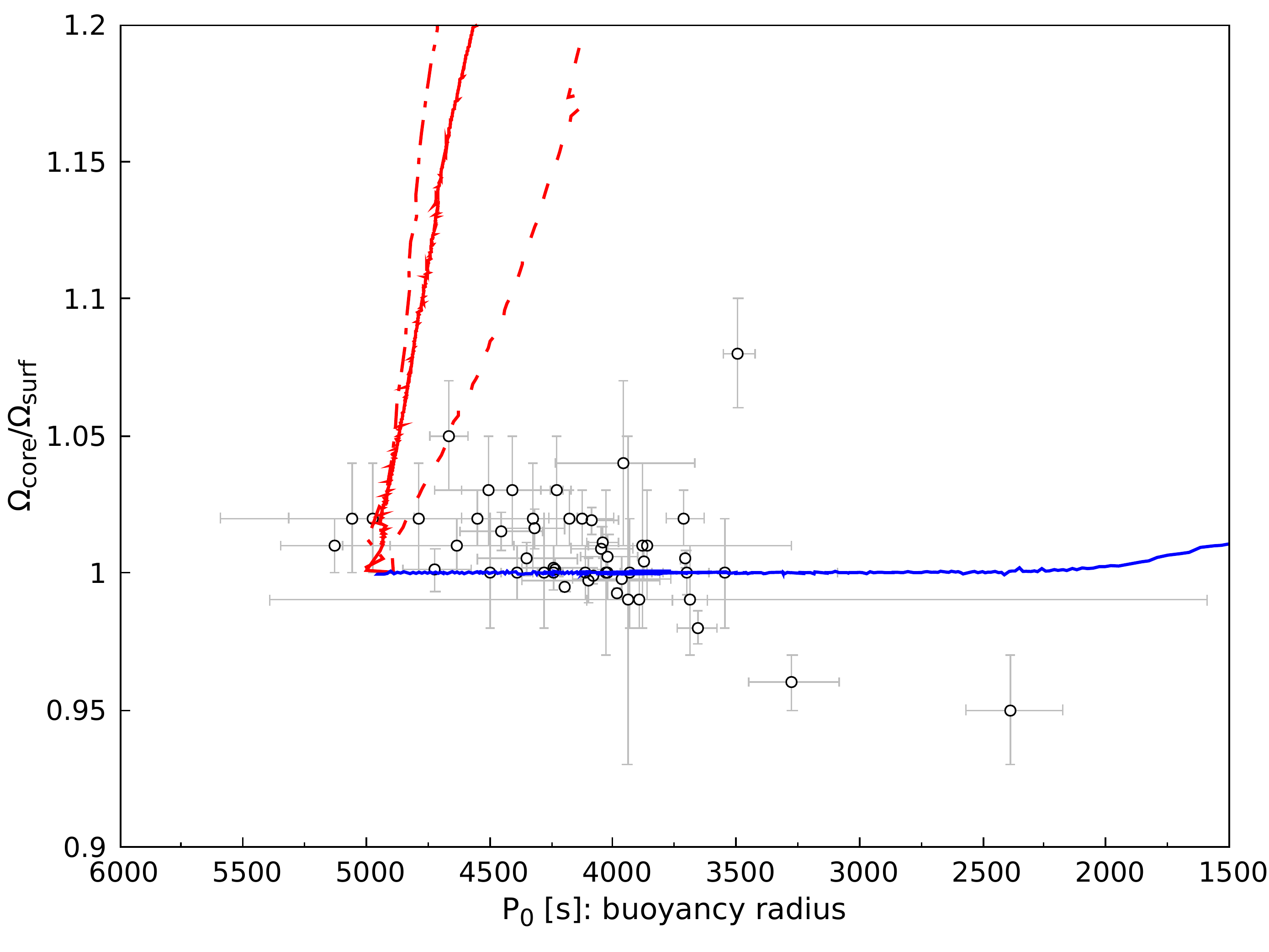}}
        \caption{Ratio of core to surface rotation rate as a function of the buoyancy radius for models with initial mass and metallicity of $M=1.5 M_{\odot}$ and $Z=0.01$, respectively.
            The data points correspond to the data presented by \citet{li20}.
            The red and blue lines are models computed with either only hydrodynamical processes (red) or both hydrodynamical processes and internal magnetic fields (blue).
            The models without internal magnetic fields are shown for three different initial rotation rates of $\Omega/2\pi= 10, 20, \text{and } 30 \mu$Hz by dashed, solid, and dot-dashed lines, respectively.
        For models with internal magnetic fields, only the one starting with $\Omega/2\pi=20 \mu$Hz is shown.}
     \label{omega_ratio_li20}
   \end{figure}

  Assuming that the modulation in the light-curves corresponds to magnetic spots at the stellar surface, we test our models with the surface rotation rates provided by \citet{li20} as a function of the buoyancy radius in Fig. \ref{omega_ratio_li20}.
  We excluded the stars labelled as eclipsing binaries in the sample of \citet{li20} for this comparison since in those cases the surface rotation rate was assumed to be equal to the orbital period of the system.
  Our `only-rotation' models quickly develop internal differential rotation in radiative regions, leading to a disagreement with data, although at lower initial rotation rates the decoupling occurs slower, as evidenced from the model shown by dashed-lines in Fig. \ref{omega_ratio_li20}.

  But in our models with internal magnetic fields the radiative regions remain strongly coupled and so the ratio of core-to-surface rotation rate is nearly unity during the whole MS, leading only to a slight decoupling as the star approaches the end of the MS as a result of the development of chemical gradients close to the convective core.
  We obtain this result in our models irrespective of the initial rotation rate assumed.
  This is why we only show the model with an initial rotation rate of $\Omega/2\pi=20 \mu$Hz in Fig. \ref{omega_ratio_li20} when magnetic fields are included.

Although models without magnetic fields are in disagreement with the data, it is still difficult to reproduce the small degree of differential rotation seen in the data with our models including internal magnetic fields.
Also, the apparently decreasing trend seen in Fig. \ref{omega_ratio_li20} remains challenging in our models with internal magnetic fields since it is not possible to generate differential rotation in a way that the surface rotates faster than the deep radiative layers through evolution because of the diffusive nature on the AM transport of the magnetic fields studied.

  \section{Discussion}
  \label{discussion}
    Overall, our work gives strong support to efficient AM transport during the MS, although its physical nature remains debatable.
    This supports the idea of at least one missing physical process able to transport AM in stellar interiors \citep{eggenberger12,marques13,ceillier13}, besides the advection of AM by meridional circulation and turbulent transport by shear instabilities.
    Previous asteroseismic results on single stars obtained by \citet{saio15} and \citet{murphy16} indicate that the AM transport in main sequence F-type stars must be highly efficient through evolution \citep[see also][]{kurtz14,hatta22}; we support this idea.
  We believe that the additional process should be active during the whole MS, leading to quasi-rigid rotation in the radiative zones, such as is the case for the Sun \citep{couvidat03}.
  We do note however that there could be still some differential rotation between the convective core and the radiative regions in $\gamma$ Dor stars, which will be the topic of a follow-up study.

 Evidence from post-MS stars such as subgiants and red giants points out to a general lack of AM transport in stellar models.
  The TS dynamo seems to satisfy well the rotational evolution of the core during the MS and evolved stages simultaneously, but there are still problems reproducing the rotation rate of subgiants and young RGB stars simultaneously \citep{eggenberger19c}.
  More constraints on the AM content in subgiants would provide crucial information to distinguish between physical scenarios; as to date this kind of constraints are available for only eight subgiants \citep{deheuvels14,deheuvels20}.
  The sample of stars studied here represent the MS counterpart of the red giants studied by \citet{gehan18} (see also \citet{mosser12}), and future studies on AM redistribution should focus on the connection between both samples and exploit other indicators such as chemical abundances.
  The present work helps us understand the physical process(es) operating in dwarfs, and in connection with the constraints from \citet{deheuvels20} for two young subgiants, it gives support to AM transport by the magnetic Tayler instability as the main candidate to resolve the AM transport problem.
  We also note that other physical processes could contribute to the transport of AM, such as internal gravity waves \citep{pincon17}, mixed modes \citep{belkacem15a,belkacem15b}, or magneto-rotational like instabilities \citep{spada16,moyano23}.

  Another way to study AM redistribution would be to use the periods of rotation of the stellar surface as given by the modulation of the light-curve, usually attributed to magnetic spots on the surface of the star.
  An issue with this approach is the possible misidentification of rotational periods as signatures of pulsations; the modulation seen in the light-curve was shown to be possibly due to g-modes excited by the convective core \citep{lee20}, and it remains a possibility despite recent advances \citep[see][]{henriksen23}.
  In that case the rotational period obtained from the modulation of the light curves would actually be the rotation rate of the convective core and although one could reach the same conclusions as we do, the hypotheses would be wrong.
  We avoid such possible misinterpretations by using the measurements of projected rotational velocities from spectroscopy.
  Although there are other processes able to broaden the spectral lines (e.g. macroturbulence, pressure broadening) they are expected to be negligible compared to the strong broadening in fast rotators; we recall that some of these stars are rotating up to 50 \% of their critical rotational velocities (see Sect. \ref{methods}).

    Moreover, it is possible to study the internal redistribution of AM through evolution with help of rotational splittings coming from hybrid pulsators with both p- and g-modes \citep[e.g.][]{dupret04,kurtz14,saio15}.
    However this is difficult to achieve for a large sample of stars since stars showing these properties are not abundant.
  In this work we show that we can avoid such inconvenient by combining data from spectroscopy for a large sample of stars and making coherent assumptions on the modelling of the AM transport.

  Future works should verify whether signatures of chemical enrichment in stellar models with strong coupling are in agreement with data.
  Also, it is necessary to verify whether our conclusions hold for more massive stars, since some targets remain open to different interpretations \citep[see][]{salmon22}.

  \section{Summary and conclusions}
  \label{conclusion}
  We investigate the efficiency of AM transport in MS stars using constraints on both the core rotation rates \citep{li20} and surface rotational velocities from spectroscopy (Sect. \ref{data}) of $\gamma$ Dor stars.
  We build upon the work of \citet{ouazzani19} and our conclusions mainly rely on the data obtained by \citet{li20} and \citet{gebruers21}.
  We show how combining information from independent techniques, we can better distinguish the different scenarios of AM transport during the MS, which otherwise remain hidden when relying on only information of the core.
  Our main results are illustrated by Figs. \ref{magvsrot_2cols}, \ref{vsini_omegac_1p5}, and \ref{omegas_omegac}.
  For the comparison with stellar models, we choose two types of models: one in which the AM transport is done only by meridional circulation and the shear instability and another one where the action of internal magnetic fields by the Tayler-Spruit dynamo is also included.
  Internal magnetic fields transport AM very efficiently, and are in better agreement with our whole analysis.

  Our strong conclusion is that the transport of AM in radiative zones during the MS of low-mass stars must be efficient.
  As a corollary, internal magnetic fields are a candidate to the missing physical process in stellar interiors.
  Our conclusion, in addition to the strong evidence from previous works for the Sun \citep{couvidat03,eggenberger22a} and evolved stars \citep[e.g.][]{mosser12}, supports the idea that an efficient process (or at least one) able to counteract the development of differential rotation in stellar interiors must act during the whole evolution of stars.
     
   
  \begin{acknowledgements}
    We thank the referee for the constructive feedback and accurate comments that helped improving the manuscript.
    FDM, PE, SJAJS, and SE have received funding from the European Research Council (ERC) under the European Union’s Horizon 2020 research and innovation programme (grant agreement No. 833925, project STAREX).
    JSGM acknowledges funding from the French National Research Agency (ANR: grant MASSIF, ANR-21-CE3-0018-02).
    FDM thanks Agostina Fil\'ocomo, Matthias Kruckow and Luca Sciarini for useful discussions.  
\end{acknowledgements}

\bibliographystyle{aa}
\bibliography{gammadors_amt}

\begin{appendix} 
  \section{Rotation profiles with strong differential rotation}
    \label{app_diffzams}
\begin{figure}[htb!]
        \resizebox{\hsize}{!}{\includegraphics{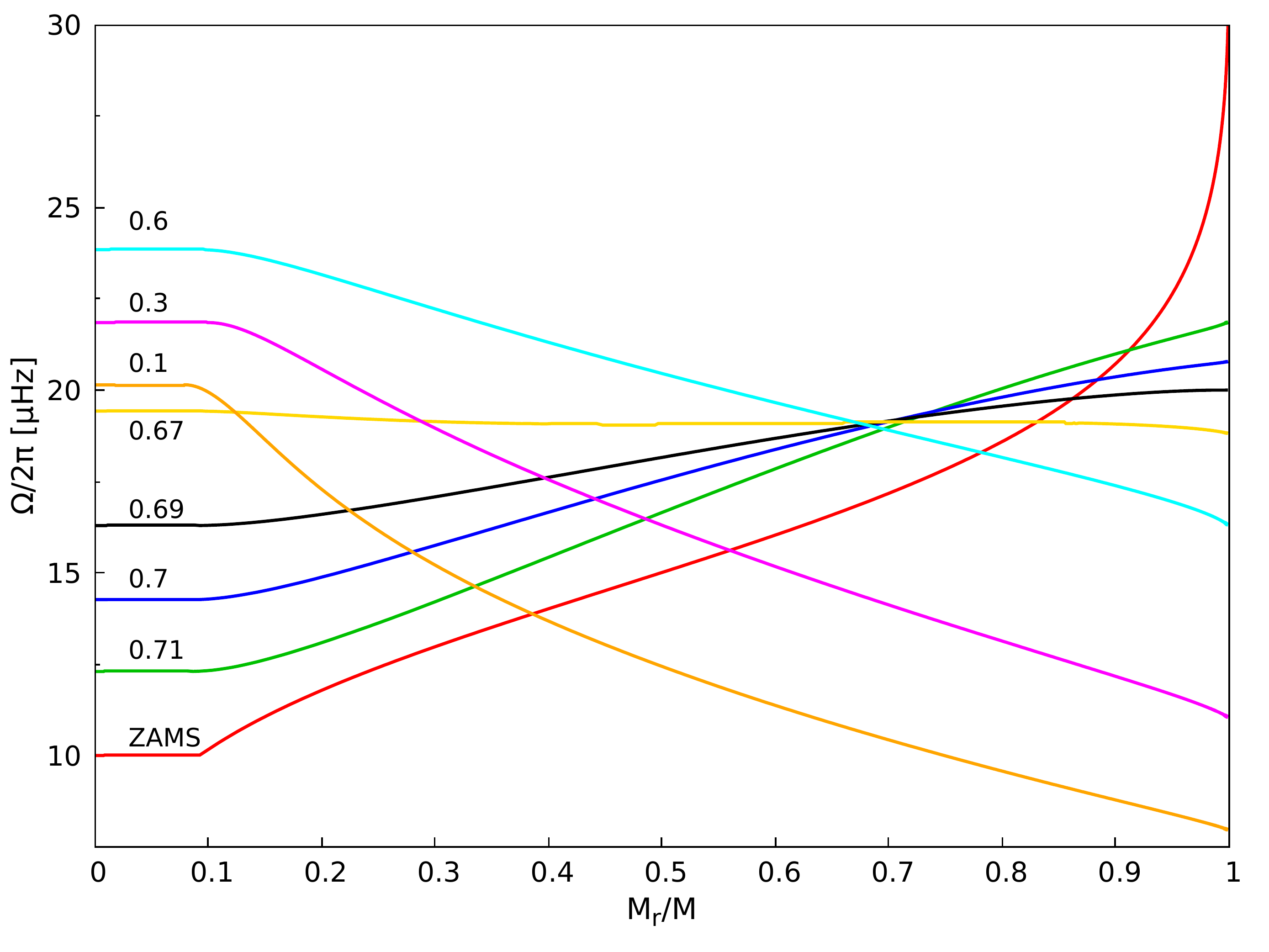}}
        \caption{Rotation rate as a function of the normalised mass-coordinate.
            The lines correspond to different evolutionary ages during the main sequence, indicated by the abundance of central hydrogen content in mass fraction over each profile.
            The red line is the initial rotation profile assumed at the ZAMS.
        }
     \label{rotprof_r03}
   \end{figure}

  In Fig. \ref{rotprof_r03} we show the evolution of the rotation profiles during the main sequence of the model presented in Fig. \ref{vsini_omegac_diffzams} with a strong rotation contrast between convective core and surface of $\Omega_{\rm surf}/\Omega_{\rm cc} = 3$ (i.e. surface rotating faster than the interior).
  The rotation profile is given as an input for the model and then only transport of angular momentum by hydrodynamical processes (meridional circulation and shear instability) is taken into account.
  The models quickly inverse the initial configuration, as soon as the central hydrogen content reaches $X_{\rm core}=0.67$ the surface starts rotating slightly slower than the core and then follows a standard evolution, as in models where solid-body rotation is assumed at the ZAMS.  

  \end{appendix}

\end{document}